\documentclass[twocolumn, twocolappendix]{aastex631}

\usepackage{natbib}
\usepackage{amsmath}
\usepackage{braket}
\usepackage{color}
\hypersetup{
 colorlinks=true,%
 linkcolor=black,
 citecolor=blue,
}

\begin{document}
\def\vec#1{\mbox{\boldmath $#1$}}
\def\rev#1{#1}
\newcommand{\average}[1]{\ensuremath{\langle#1\rangle} }

\title{\large{\bf Formulating Mass-Loss Rates for Sun-like Stars: A Hybrid Model Approach
}}

	\author{Munehito Shoda}
	\affiliation{Department of Earth and Planetary Science, School of Science, The University of Tokyo, 7-3-1 Hongo, Bunkyo-ku, Tokyo, 113-0033, Japan}
	\author{Steven R. Cranmer}
	\affiliation{Department of Astrophysical and Planetary Sciences, Laboratory for Atmospheric and Space Physics, University of Colorado, Boulder, CO 80309, USA}
	\author{Shin Toriumi}
	\affiliation{Institute of Space and Astronautical Science, Japan Aerospace Exploration Agency, 3-1-1 Yoshinodai, Chuo-ku, Sagamihara, Kanagawa 252-5210, Japan}

	\correspondingauthor{Munehito Shoda}
	\email{shoda.m.astroph@gmail.com}

\begin{abstract}
    We observe an enhanced stellar wind mass-loss rate from low-mass stars exhibiting higher X-ray flux. 
    This trend, however, does not align with the Sun, where no evident correlation between X-ray flux and mass-loss rate is present. 
    To reconcile these observations, we propose a hybrid model for the stellar wind from solar-type stars, incorporating both Alfv\'en wave dynamics and flux emergence-driven interchange reconnection, an increasingly studied concept guided by the latest heliospheric observations.
    For establishing a mass-loss rate scaling law, we perform a series of magnetohydrodynamic simulations across varied magnetic activities.
    Through a parameter survey concerning the surface (unsigned) magnetic flux ($\Phi^{\rm surf}$) and the open-to-surface magnetic flux ratio ($\xi^{\rm open} = \Phi^{\rm open}/\Phi^{\rm surf}$), we derive a scaling law of the mass-loss rate given by $\dot{M}_w/\dot{M}_{w,\odot} = \left( \Phi^{\rm surf} / \Phi^{\rm surf}_\odot \right)^{0.52}\left( \xi^{\rm open} / \xi^{\rm open}_\odot \right)^{0.86}$, where $\dot{M}_{w,\odot} = 2.0 \times 10^{-14} \ M_\odot {\rm \ yr}^{-1}$, $\Phi^{\rm surf}_\odot = 3.0 \times 10^{23} {\rm \ Mx}$, and $\xi^{\rm open}_\odot = 0.2$. 
    By comparing cases with and without flux emergence, we find that the increase in the mass-loss rate with the surface magnetic flux can be attributed to the influence of flux emergence.
    Our scaling law demonstrates an agreement with solar wind observations spanning 40 years, exhibiting superior performance when compared to X-ray-based estimations.
    Our findings suggest that flux emergence may play a significant role in the stellar winds of low-mass stars, particularly those originating from magnetically active stars.
    
\end{abstract}


\section{Introduction} \label{sec:introduction}

Low-mass main-sequence stars exhibit intrinsic magnetic fields \citep{Saar_1996_proceedings, Saar_2001_proceedings, Reiners_2009_ApJ, Vidotto_2014_MNRAS, Kochukhov_2020_AA, Reiners_2022_AA} generated through the dynamo process operating within the convection zone \citep{Parker_1955_ApJ, Brun_2004_ApJ, Hotta_2016_Science}.
The stellar magnetic field drives various physical phenomena. 
Upward energy transport of surface convection energy via the magnetic field \citep{Alfven_1947_MNRAS, Osterbrock_1961_ApJ, Parker_1983_ApJ} results in coronal heating and high-energy photon emission \citep{Vaiana_1981_ApJ, Pallavicini_1981_ApJ, Pizzolato_2003_AA, Sanz-Forcada_2011_AA, Wright_2016_Nature, Magaudda_2020_AA}. 
Magnetic reconnection of large-scale magnetic fields triggers stellar flares \citep{Masuda_1994_Nature, Shibata_2011_LRSP, Namekata_2017_ApJ} and the production of energetic particles \citep{Reames_1999_SSRev}. 
Furthermore, the stellar magnetic field drives the formation of outflows, both in a quasi-steady \citep[stellar wind,][]{Parker_1958_ApJ, Kopp_1976_SolPhys, Velli_1994_ApJ} and transient manner \citep[coronal mass ejection,][]{Antiochos_1999_ApJ_A_model_for_solar_coronal_mass_ejection, Cranmer_2017_ApJ, Veronig_2021_NatAs, Namekata_2021_NatAs}. 
These stellar magnetic activities are recognized as crucial factors in regulating the evolution of stellar systems and are extensively studied in the contexts of both stellar and planetary sciences.

The primary focus of this study is the physical properties of the stellar wind, which play significant roles in the evolution of the stellar system as follows.
The magnetized stellar wind transports angular momentum outward through magnetic braking \citep{Kraft_1967_ApJ, Weber_1967_ApJ, Mestel_1968_MNRAS, Sakurai_1985_AA, Kawaler_1988_ApJ}, resulting in the slowing down of the stellar rotation \citep{Skumanich_1972_ApJ, Barnes_2003_ApJ, Irwin_2009_proceedings,van_Saders_2016_Nature}.
The stellar wind also influences planetary evolution due to its non-negligible impact on planetary outflows \citep{Bourrier_2013_AA, Kislyakova_2013_AsBio, Cohen_2015_ApJ, Bourrier_2016_AA, Kislyakova_2019_AA, McCann_2019_ApJ, Vidotto_2020_MNRAS, Mitani_2022_MNRAS}.
It has been suggested that a portion of Earth's water may have originated from the solar wind \citep{Daly_2021_NatAs}.
To quantitatively understand these processes, the mass-loss rate of the stellar wind (denoted as $\dot{M}_w$) is an essential parameter. Consequently, the scaling law of $\dot{M}_w$ serves as a crucial building block in the study of the stellar wind and its effects on the evolution of the whole stellar system.

The difficulty in measuring mass-loss rates has resulted in a limited number of observed values, 
which are indirectly obtained through methods such as astrospheric Lyman-alpha absorption \citep{Wood_2002_ApJ, Wood_2005_ApJ, Wood_2014_ApJ, Wood_2021_ApJ}, slingshot prominences \citep{Jardine_2019_MNRAS, Jardine_2020_MNRAS}, and planetary responses to the stellar wind \citep{Vidotto_2017_MNRAS, Kavanagh_2019_MNRAS}. 
It is important to note that direct observations using radio waves \citep{Lim_1996_ApJ_outflow_from_late_type_stars, Lim_1996_ApJ_Mdot_of_ProxCen, Gaidos_2000_GRL, Fichtinger_2017_AA, Vidotto_2017_AA} and X-rays \citep{Wargelin_2002_ApJ} provide only the upper limit of $\dot{M}_w$.
The observed mass-loss rate roughly exhibits a power-law dependence on the X-ray flux \citep[measured in the ROSAT band,][]{Wood_2021_ApJ, Vidotto_2021_LRSP}. 
However, the mass-loss rate of the solar wind remains nearly constant over the activity cycle, despite a factor of 20 variation in the X-ray flux \citep{Johnstone_2015_AA}. 
This discrepancy between solar and stellar observations of $\dot{M}_w$ presents a major concern when employing the observational scaling laws.

The purpose of this work is to establish a theoretical scaling law for $\dot{M}_w$ \citep{Schroder_2005_ApJ, Cranmer_2011_ApJ, Suzuki_2013_PASJ, Suzuki_2018_PASJ, Chebly_2023_MNRAS} that aligns with both solar and stellar observations. 
Solar wind models generally rely on either Alfv\'en-wave acceleration \citep{Ofman_1995_JGR, Ofman_1998_JGR, Matthaeus_1999_ApJ, Dmitruk_2002_ApJ, Ofman_2004_JGRA, Suzuki_2005_ApJ, Verdini_2007_ApJ, Verdini_2010_ApJ, Perez_2013_ApJ, van_der_Holst_2014_ApJ, van_Ballegooijen_2016_ApJ, Usmanov_2018_ApJ, Shoda_2019_ApJ, Reville_2020_ApJ, Matsumoto_2021_MNRAS, Magyar_2021_ApJ_GPM_solar_wind, Schleich_2023_AA} or loop-opening acceleration \citep{Fisk_1999_JGR, Fisk_2003_JGR, Rappazzo_2012_ApJ, Lionello_2016_ApJ, Bale_2023_Nature, Drake_2023_arXiv, Kumar_2023_ApJ, Raouafi_2023_ApJ, Chitta_2023_NatureAstronomy, Chitta_2023_arXiv}.
Recent observations of magnetic switchbacks \citep{Bale_2019_Nature, Kasper_2019_Nature, Bale_2021_ApJ, Telloni_2022_ApJ} appear to support the loop-opening scenario \citep{Fisk_2020_ApJ, Tenerani_2020_ApJS, Zank_2020_ApJ, Drake_2021_AA, Magyar_2021_ApJ_switchback_II}, though alternative or hybrid models for switchback generation have been put forward \citep{Squire_2020_ApJ, Schwadron_2021_ApJ, Shoda_2021_ApJ, Wyper_2022_ApJ}.
It has been proposed that a hybrid model (incorporating both Alfv\'en wave and loop-opening effects) is required to accurately model the solar wind \citep{Wang_2020_ApJ, Iijima_2023_ApJ}.
With this in mind, we aim to derive the scaling law of $\dot{M}_w$ based on the hybrid model of solar (and stellar) wind and assess its compatibility with solar and stellar observations.

The rest of this paper is organized as follows.
Section~\ref{sec:model} provides a comprehensive description of the proposed hybrid model, detailing the key aspects and components that contribute to its development.
In Section~\ref{sec:result}, we present the simulation results, illustrating the performance of the hybrid model and its agreement with solar and stellar observations.
Finally, Section~\ref{sec:summary_discussion} offers a summary of the paper, along with a discussion of potential applications and limitations of the study, highlighting areas for future research and improvement.

\section{Model \label{sec:model}} 

\subsection{Model overview and assumption}

In this study, we focus on solar-type stars, namely stars possessing the same luminosity, mass, radius, and metallicity as the Sun. 
The only stellar parameter that we consider to exhibit variability is the rotation period. 
\rev{Altering stellar rotation has two effects. 
First, wind characteristics change due to centrifugal force when equatorial rotation exceeds a small fraction of the breakup speed \citep{Matt_2012_ApJ}. 
Second, the rotation period directly impacts the stellar dynamo, affecting the properties of the stellar magnetic field \citep{Saar_1996_proceedings, Saar_2001_proceedings, Reiners_2009_ApJ, See_2019_ApJ, Reiners_2022_AA}. 
In this study, we solely consider changes in the stellar magnetic field for simplicity.
Our assumption is validated for rotation rates less than ten times solar, or equivalently, in the unsaturated regime in terms of X-ray emission \citep{Pizzolato_2003_AA, Wright_2016_Nature}.}

We adopt the assumption that non-magnetic photospheric parameters -- temperature, density, scale height, and correlation length -- align with those of the Sun, as detailed in Table~\ref{table:photospheric_parameters}. 
The rationale is that non-magnetic processes (nuclear reaction, radiative transfer, and convection) presumably shape stellar internal structures \citep{Paxton_2011_ApJS}. 
Nonetheless, the potential influence of magnetic fields on the photosphere deserves further investigation in the future.

In this work, $B_{r,\ast}$ symbolizes the field intensity of localized magnetic patches located within intergranular lanes.
Following \cite{Cranmer_2011_ApJ}, we assume that these fields exhibit a magnetic pressure equivalent to the surrounding gas pressure, referred to as the equipartition field. Specifically,
\begin{align}
    B_{r,\ast} = 1340 {\rm \ G},
\end{align}
an assertion consistent with solar observations \citep{Tsuneta_2008_ApJ}.
Generally, the equipartition field substantially surpasses the surface-averaged magnetic field in magnitude,
and the proportion between these two field intensities is often interpreted as the filling factor of the magnetic field on the surface \citep{Saar_1996_proceedings, Saar_2001_proceedings, Cranmer_2017_ApJ}.
Although the assumption of equipartition is commonly used in the literature, it is important to note that the magnetic field could be larger than the equipartition on the photosphere \citep{Okamoto_2018_ApJ, Kochukhov_2020_AA}.
We also note that the employed value of $B_{r,\ast}$ is slightly larger than the typical field strength of emerging flux \citep{Centeno_2007_ApJ}.

\renewcommand{\arraystretch}{1.5}
\begin{table}[t!]
\centering
  \begin{tabular}{p{3.5em} p{10em} p{8.5em}}
    symbol
    & definition
    & value
    \\ \hline \hline
    $r_\ast$
    & radial distance
    & $6.96 \times 10^{10} {\rm \ cm}$\\
    $T_\ast$
    & (effective) temperature
    & $5.78 \times 10^3  {\rm \ K}$\\
    $\rho_\ast$
    & mass density 
    & $1.88 \times 10^{-7} {\rm \ g \ cm^{-3}}$\\
    $H_\ast$
    & pressure scale height
    & $1.74 \times 10^7 {\rm \ cm}$\\
    $\lambda_{\perp,\ast}$
    & correlation length
    & $2.00 \times 10^8 {\rm \ cm}$\\
    \hline \hline
  \end{tabular}
  \vspace{0.5em}
  \caption{Surface parameters employed in our model. 
  }
  \label{table:photospheric_parameters}
\end{table}

\vspace{1em}
\subsection{Basic equations \label{sec:basic_equations}}

In this study, we solve the MHD equations in a one-dimensional system along a super-radially expanding flux tube. 
For simplicity, we assume the flux tube to be vertically aligned. 
Let $r$ be the coordinate along the axis of the flux tube and $x$ and $y$ be the transverse components. 
The basic equations are then written as follows \citep{Shoda_2021_AA}.
\begin{gather}
    \frac{\partial \boldsymbol{U}}{\partial t} + \frac{1}{r^2 f^{\rm open}} \frac{\partial}{\partial r} \left( \boldsymbol{F} r^2 f^{\rm open} \right) = \boldsymbol{S},
    \label{eq:basic_conservation_law}
\end{gather}
\begin{gather}
    \vec{U} =
    \left(
    \begin{array}{c}
    \rho \\
    \rho v_r \\
    \rho v_x \\
    \rho v_y \\
    B_x \\
    B_y \\
    e
    \end{array}
    \right), \hspace{0.5em}
    \vec{F} =
    \left(
    \begin{array}{c}
    \rho v_r \\
    \rho v_r^2 + p_{\rm T} \\
    \rho v_r v_x - \dfrac{B_r B_x}{4\pi} \\
    \rho v_r v_y - \dfrac{B_r B_y}{4\pi} \\
    v_r B_x - v_x B_r \\
    v_r B_y - v_y B_r \\
    \left( e + p_{\rm T} \right) v_r - B_r \dfrac{\vec{v}_\perp \cdot \vec{B}_\perp}{4\pi}
    \end{array}
    \right), \label{eq:flux_terms}
\end{gather}
\begin{gather}
    \vec{S}
    =
    \left(
    \begin{array}{c}
    0 \\[4pt]
    \left( p +\dfrac{1}{2} \rho \vec{v}_\perp^2 \right) \dfrac{d}{dr} \ln \left(r^2 f^{\rm open}\right) - \rho \dfrac{GM_\ast}{r^2} \\[8pt]
    \dfrac{1}{2} \left( - \rho v_r v_x + \dfrac{B_r B_x}{4\pi} \right) \dfrac{d}{dr} \ln \left(r^2 f^{\rm open}\right)  + \rho D^{\rm v}_x \\[8pt]
    \dfrac{1}{2} \left( - \rho v_r v_y + \dfrac{B_r B_y}{4\pi} \right) \dfrac{d}{dr} \ln \left(r^2 f^{\rm open}\right)  + \rho D^{\rm v}_y \\[8pt]
    \dfrac{1}{2} \left( v_r B_x - v_x B_r \right) \dfrac{d}{dr} \ln \left(r^2 f^{\rm open}\right) + \sqrt{4 \pi \rho} D^{\rm b}_x\\[8pt]
    \dfrac{1}{2} \left( v_r B_y - v_y B_r \right) \dfrac{d}{dr} \ln \left(r^2 f^{\rm open}\right) + \sqrt{4 \pi \rho} D^{\rm b}_y\\[8pt]
    - \rho v_r \dfrac{GM_\ast}{r^2} + Q^{\rm cnd} + Q^{\rm rad} + Q^{\rm FE}
    \end{array} 
    \right), \label{eq:source_terms}
\end{gather}
where
\begin{align}
    &\boldsymbol{v}_\perp = v_x \boldsymbol{e}_x + v_y \boldsymbol{e}_y, \hspace{1em}
    \boldsymbol{B}_\perp = B_x \boldsymbol{e}_x + B_y \boldsymbol{e}_y,
\end{align}
and the total pressure $p_{\rm T}$ and total energy density $e$ are expressed as
\begin{align}
    &p_{\rm T} = p + \frac{\vec{B}_\perp^2}{8\pi}, \hspace{2em} e = e^{\rm int} + \frac{1}{2} \rho \vec{v}^2 + \frac{\vec{B}_\perp^2}{8\pi},
\end{align}
where $e^{\rm int}$ is the internal energy density per unit volume. 
We employ the equation of state for partially ionized hydrogen gas to determine $e^{\rm int}$ as a function of $p$ and $\rho$ \citep{Vogler_2005_AA, Shoda_2021_AA}.
$f^{\rm open}$ denotes the filling factor of the open magnetic flux.
The terms $D^{\rm v}_{x,y}$ and $D^{\rm b}_{x,y}$ represent the phenomenological terms of turbulent dissipation.
The heating rates by thermal conduction and radiation are denoted by $Q^{\rm cnd}$ and $Q^{\rm rad}$, respectively. 
The heating rate of interchange reconnection driven by flux emergence is represented by $Q^{\rm FE}$.

We assert that although turbulent dissipation components do not feature in the energy equation, our model accounts for turbulent heating. This is clarified by elucidating the conservation law of mechanical energy (\rev{$e^{\rm mech} = e - e^{\rm int}$}), as shown below:
\begin{align}
    \frac{\partial}{\partial t} e^{\rm mech} &+ \nabla \cdot \left[ \left( e^{\rm mech} + p_{\rm T} \right) \boldsymbol{v} - \frac{\boldsymbol{B}}{4 \pi} \left( \boldsymbol{v}_\perp \cdot \boldsymbol{B}_\perp \right) \right] \notag \\ &= p \nabla \cdot \boldsymbol{v} - Q^{\rm turb} - \rho v_r \frac{GM_\ast}{r^2}, \label{eq:mechanical_energy_conservation}
\end{align}
where
\begin{align}
    Q^{\rm turb} = - \rho \left( v_x D_x^{\rm v} + v_y D_y^{\rm v} + \frac{B_x}{\sqrt{4 \pi \rho}} D_x^{\rm b} + \frac{B_y}{\sqrt{4 \pi \rho}} D_y^{\rm b} \right)
    \label{eq:heating_rate_awt}
\end{align}
represents the turbulent dissipation rate of Alfv\'en wave. 
We note that the divergence of a vector field $\boldsymbol{X}$ is given by
\begin{align}
    \nabla \cdot \boldsymbol{X} = \frac{1}{r^2 f^{\rm open}} \frac{\partial}{\partial r} \left( X_r r^2 f^{\rm open} \right),
\end{align}
where $X_r$ is the $r$ component of $\boldsymbol{X}$.

The internal energy equation is obtained by subtracting Eq.\eqref{eq:mechanical_energy_conservation} from the equation for $e$. 
The result is
\begin{align}
    \frac{\partial}{\partial t} e^{\rm int} &+ \nabla \cdot \left( e^{\rm int} \boldsymbol{v} \right) \notag \\
    &= - p \nabla \cdot \boldsymbol{v} + Q^{\rm turb} + Q^{\rm cnd} + Q^{\rm rad} + Q^{\rm FE}.
\end{align}
This equation includes the turbulent dissipation rate as a heating term. 
Thus, by solving Eq.s~\eqref{eq:basic_conservation_law}--\eqref{eq:source_terms}, the heating effect of turbulent dissipation is fully incorporated. 
The same arguments holds for numerical dissipation and heating.

\begin{figure}[t!]
\centering
\includegraphics[width=75mm]{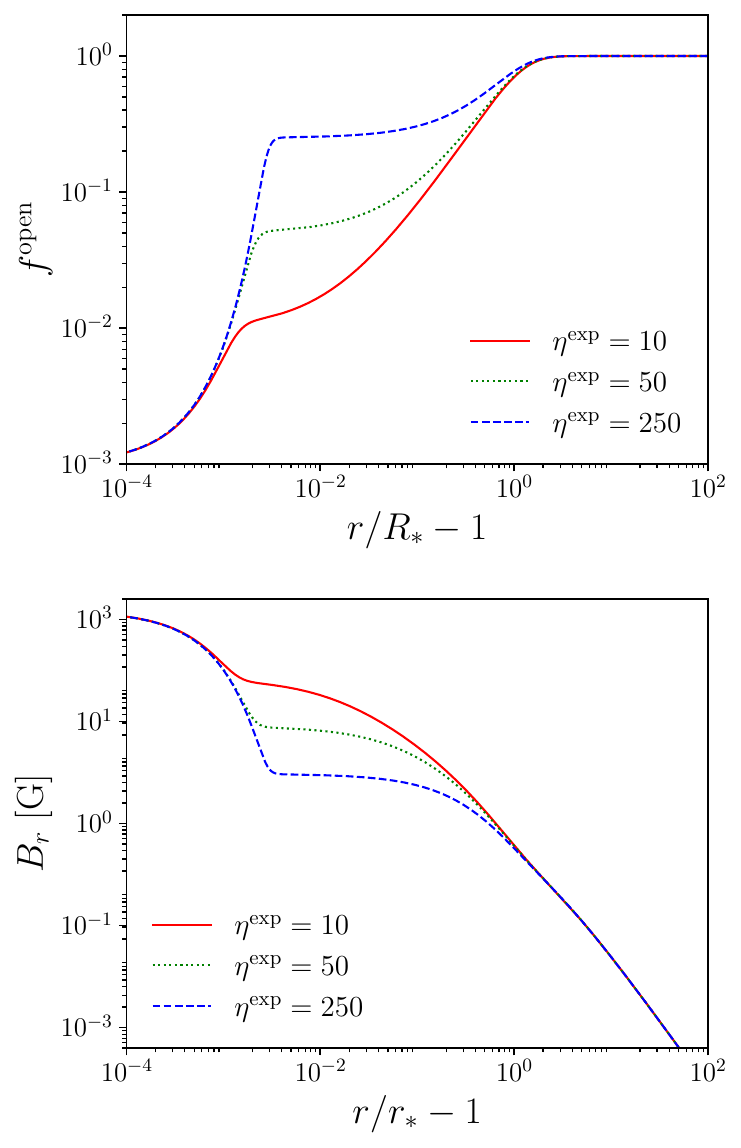}
\caption{
Radial profiles of the open-flux filling factor ($f^{\rm open}$, top) and radial magnetic field ($B_r$, bottom) as defined by Eq.s~\eqref{eq:fopen}-\eqref{eq:br_radial_profile}. 
Three lines represent varying chromospheric expansion parameters: $\eta^{\rm exp} = 10$ (red solid line), $\eta^{\rm exp} = 50$ (green dotted line), and $\eta^{\rm exp} = 250$ (blue dashed line). 
$f^{\rm open}_\ast$ and $B_{r,\ast}$ are fixed to $1.00 \times 10^{-3}$ and $1.34 \times 10^3$ G, respectively.
}
\label{fig:filling_factor_example}
\vspace{1em}
\end{figure}

\vspace{1em}
\subsection{Flux-tube parameters \label{sec:flux_tube}}

$f^{\rm open}$ regulates the shape of the open field lines and is in reality determined by the MHD force balance. 
However, since our model does not account for force balance in the cross-field direction, we need to prescribe $f^{\rm open}$ as a function of $r$. 
In this study, we formulate $f^{\rm open}(r)$ using its surface value $f^{\rm open}_\ast$ and the flux-tube expansion parameter ($\eta^{\rm exp}$) as follows
\begin{align}
    & f^{\rm open} (r) =  f^{\rm open}_\ast \frac{f_1 (r) f_2 (r)}{\sqrt{f_1 (r)^2 + f_2 (r)^2}}, \label{eq:fopen} \\
    & f_1 (r) = \frac{\eta^{\rm exp}}{\sqrt{\left( \eta^{\rm exp} \right)^2 -1 }}\exp \left( \frac{r-r_\ast}{H^{\rm exp}} \right), \\
    & f_2 (r) = \frac{\mathcal{F}(r) + \eta^{\rm exp} f^{\rm open}_\ast + \mathcal{F}(r_\ast) \left(  \eta^{\rm exp} f^{\rm open}_\ast -1 \right)}{f^{\rm open}_\ast \left(\mathcal{F}(r)+1\right) },
\end{align}
where
\begin{align}
    \mathcal{F}(r) = \exp \left( \frac{r/r_\ast - 1.3}{0.5} \right).
\end{align}
The functions $f_1 (r)$ and $f_2 (r)$ represent the flux-tube expansion in the chromosphere \citep{Ishikawa_2021_ScienceAdvances} and corona \citep{Kopp_1976_SolPhys}, respectively. 
The parameter $\eta^{\rm exp}$ represents the flux-tube expansion factor in the chromosphere. The expansion length scale is set to $H^{\rm exp} = 2H_\ast$, where $H_\ast$ is the pressure scale height on the photosphere.

$f^{\rm open}_\ast$ and $\eta^{\rm exp}$ are connected to the unsigned surface magnetic flux ($\Phi^{\rm surf}$) and open magnetic flux ($\Phi^{\rm open}$) through magnetic-flux conservation.
\begin{align}
    \eta^{\rm exp} = \frac{4 \pi r_\ast^2 B_{r,\ast}}{\Phi^{\rm surf}}, \ \ \ \ f^{\rm open}_\ast = \frac{\Phi^{\rm open}}{4 \pi r_\ast^2 B_{r,\ast}}, \label{eq:eta_fopen}
\end{align}
where $\Phi^{\rm surf}$ and $\Phi^{\rm open}$ represent the unsigned surface and open magnetic fluxes, respectively. 
We infer $\eta^{\rm exp}$, presuming equivalence between the unsigned magnetic flux at the photospheric and coronal base levels. 
This implies the absence of magnetic loops closing within the photosphere or chromosphere--an assumption that may overstate reality given that a subset of these loops peak at heights less than $1000$ km \citep{Wiegelmann_2004_SolPhys, Cranmer_2010_ApJ_can_the_solar_wind}.
The product $\eta^{\rm exp} f^{\rm open}_\ast$ signifies the ratio of open to surface magnetic flux and is denoted as $\xi^{\rm open}$:
\begin{align}
    \xi^{\rm open} = \eta^{\rm exp} f^{\rm open}_\ast = \Phi^{\rm open} / \Phi^{\rm surf}. \label{eq:xiopen}
\end{align}

With the formulation of $f^{\rm open} (r)$, we can deduce the radial component of the magnetic field $B_r$ from the magnetic-flux conservation of the flux tube:
\begin{align}
    B_r (r) = B_{r,\ast} \frac{r_\ast^2 f^{\rm open}_\ast}{r^2 f^{\rm open} (r)}. \label{eq:br_radial_profile}
\end{align}

\vspace{1em}
\subsection{Alfv\'en-wave turbulence \label{sec:awt}}

Alfv\'en-wave turbulence is likely responsible for converting wave energy into thermal energy. 
Since Alfv\'en-wave turbulence is fundamentally a multi-dimensional process, with energy cascading primarily in the cross-field direction, the effect of turbulent dissipation must be phenomenologically incorporated in one-dimensional models. 
In this work, we employ the conventional one-point closure model for the phenomenology of Alfv\'en-wave turbulence \citep{Hossain_1995_PhysFluids, Matthaeus_1999_ApJ, Dmitruk_2002_ApJ}. 
It is important to note, however, that there are improved models available, such as the nearly-incompressible MHD turbulence model and the MHD shell model. 

Following \citet{Shoda_2018_ApJ_a_self-consistent_model}, we express $D^{\rm v}_{x,y}$ and $D^{\rm b}_{x,y}$ in terms of Els\"asser variables \citep{Elsasser_1950_PR} as follows:
\begin{align}
    D^{\rm v}_{x,y} &= - \frac{c_d}{4\lambda_\perp} \left( \left| z_{x,y}^+ \right| z_{x,y}^- + \left| z_{x,y}^- \right| z_{x,y}^+ \right), \label{eq:phenomenological_awt_vsource} \\
    D^{\rm b}_{x,y} &= - \frac{c_d}{4\lambda_\perp} \left( \left| z_{x,y}^+ \right| z_{x,y}^- - \left| z_{x,y}^- \right| z_{x,y}^+ \right), \label{eq:phenomenological_awt_bsource}
\end{align}
where
\begin{align}
    \boldsymbol{z}_\perp^\pm = \boldsymbol{v}_\perp \mp \boldsymbol{B}_\perp/\sqrt{4 \pi \rho},
\end{align}
$c_d$ is a dimensionless coefficient, and $\lambda_\perp$ is the perpendicular correlation length of turbulence. 
Previous three-dimensional reduced MHD simulations \citep{Perez_2013_ApJ, van_Ballegooijen_2016_ApJ, van_Ballegooijen_2017_ApJ, Chandran_2019_JPP} and shell-model calculations \citep{Verdini_2019_SolPhys} have suggested that $c_d$ is significantly smaller than unity. 
The inferred best-fit parameter is $c_d = 0.1$ \citep{van_Ballegooijen_2016_ApJ}, which is used in this work.

In this study, instead of solving the radial transport of the correlation length \citep{Matthaeus_1994_ApJ, Breech_2008_JGR, Cranmer_2012_ApJ, Zank_2018_ApJ}, we prescribe $\lambda_\perp$ as a function of $r$. 
It is often assumed that $\lambda_\perp$ scales with the radius of the flux tube \citep{Hollweg_1986_JGR, Verdini_2007_ApJ, van_der_Holst_2014_ApJ}, which is likely satisfied in the magnetically-dominated region.
However, in the lower atmosphere, the spatial scale of fluctuation is possibly regulated by granulation rather than the magnetic field expansion. 
Indeed, the granular pattern of velocity appears to persist up to the low chromosphere \citep{Salucci_1994_AA, Espagnet_1995_AA, Kostik_2009_AA}. 
Based on these arguments, we assume that $\lambda_\perp$ undergoes radial expansion until the flux tube completes its expansion in the chromosphere, and thereafter follows field-aligned expansion. 
This can be expressed as:
\begin{align}
    \lambda_\perp (r) = \lambda_{\perp,\ast} \frac{r}{r_\ast} \max \left[ 1, \sqrt{f^{\rm open}(r) \eta^{\rm exp}} \right],
\end{align}
where the correlation length in the photosphere is set as $\lambda_{\perp, \ast} = 2000$ km (Table~\ref{table:photospheric_parameters}).

We utilize a larger $\lambda_{\perp,\ast}$ value than commonly applied in past studies \citep[$\sim 150$ km,][]{Cranmer_2011_ApJ, Shoda_2020_ApJ, Shoda_2021_AA}. This is attributed to the energy-containing scale in the horizontal velocity field of the photosphere resembling the granular scale ($\sim 1000$ km) \citep{Matsumoto_2010_ApJ, Ishiwaka_2022_AA}. Given that larger convective structures persist to higher altitudes \citep{Kostik_2009_AA}, the granular scale is expected to be greater in the upper atmosphere. Consequently, we adopt $\lambda_{\perp,\ast} = 2000$ km, slightly exceeding the typical granular scale ($1000$ km). The legitimacy of this presumption hinges on the wave generation mechanism; if Alfv\'en waves originate from inter-granular-scale vortices \citep{van_Ballegooijen_2011_ApJ, Finley_2022_AA, Kuniyoshi_2023_ApJ, Breu_2023_arXiv}, the scale of intergranular lane should be designated as \rev{$\lambda_{\perp,\ast} = 100-200 {\rm \ km}$ \citep{Berger_2001_ApJ, van_Ballegooijen_2011_ApJ}}.
\rev{In future studies, the radial profile of $\lambda_\perp$ should be more accurately prescribed to align with observations \citep{Abramenko_2013_ApJ, Sharma_2023_NatAs}.}

\vspace{1em}
\subsection{Thermal conduction and radiation}

The heat generated due to thermal conduction can be expressed in terms of the heat flux, denoted by $F^{\rm cnd}$, as follows:
\begin{align}
    Q^{\rm cnd} = - \frac{1}{r^2 f^{\rm open}} \frac{\partial}{\partial r} \left( F^{\rm cnd} r^2 f^{\rm open} \right).
    \label{eq:heating_rate_conduction}
\end{align}
For the calculation of $F^{\rm cnd}$, we utilize the Spitzer-H\"arm type flux \citep{Spitzer_1953_PhysRev} incorporating quenching in low-density regions \citep{Shoda_2020_ApJ}:
\begin{align}
    F^{\rm cnd} = - \kappa_{\rm SH} T^{5/2} \frac{\partial T}{\partial r} \min \left[1,\frac{\rho}{\rho^{\rm cnd}} \right].
\end{align}

Introducing quenching enables us to mitigate the stringent time-step constraint arising from thermal conduction. 
However, it is crucial to ensure that conduction is not quenched within the middle corona, where it significantly influences the energy balance. 
In this study, we use $\rho^{\rm cnd} = 10^{-20}$ g cm$^{-3}$, \rev{leading to conduction quenching above $r/R_\odot \approx 10-20$, where the wind is supersonic}.
This results in a considerably reduced conductive flux compared to observations at 1 au \citep{Salem_2003_ApJ, Bale_2013_ApJ}; 
nevertheless, it is not expected to substantially impact the mass-loss rate, as quenching starts to take effect in the supersonic region.

The radiative \rev{heating} term, $Q^{\mathrm{rad}}$, consists of two components: the optically thick ($Q^{\mathrm{thck}}$) and optically thin ($Q^{\mathrm{thin}}$) contributions. These are combined as follows \citep{Shoda_2021_AA}:
\begin{gather}
    Q^{\mathrm{rad}} = (1-\xi^{\mathrm{rad}}) Q^{\mathrm{thck}} + \xi^{\mathrm{rad}} Q^{\mathrm{thin}}, \notag \\
    \xi^{\mathrm{rad}} = {\mathrm{min}} \left[ 1, \exp \left( - \frac{p}{p^{\mathrm{rad}}} \right) \right],
    \label{eq:heating_rate_radiation}
\end{gather}
where the parameter $p^{\mathrm{rad}}/p_\ast$ is set to $0.1$.

$Q^{\mathrm{thck}}$ is approximated by an exponential cooling process \citep{Gudiksen_2005_ApJ}, which drives the internal energy toward a reference value on a specified timescale:
\begin{align}
    Q^{\mathrm{thck}} = - \frac{1}{\tau^{\mathrm{thck}}} \left( e^{\mathrm{int}} - e^{\mathrm{int}}_{\mathrm{ref}} \right),
\end{align}
where the timescale $\tau^{\mathrm{thck}}$ is defined as:
\begin{align}
    \tau^{\mathrm{thck}} = 0.1 \ {\mathrm{s}} \left( \frac{\rho}{\rho_\ast} \right)^{-1/2},
\end{align}
and $e^{\mathrm{int}}_{\mathrm{ref}}$ denotes the reference internal energy, which corresponds to a temperature equivalent to the surface temperature. Consequently, the optically thick cooling mechanism serves to bring the atmospheric temperature closer to the surface value.

The optically-thin cooling rate, denoted as $Q^{\mathrm{thin}}$, is expressed in terms of the radiative loss function ($\Lambda$) as follows:
\begin{align}
    Q^{\mathrm{thin}} = - n_{\mathrm{H}} n_e \Lambda (T),
\end{align}
where $n_{\mathrm{H}}$ and $n_e$ represent the number densities of hydrogen atoms and electrons, respectively. 
The function $\Lambda (T)$ is computed using the CHIANTI atomic database version 7 \citep{Dere_1997_AA, Landi_2012_ApJ} with the photospheric abundance and extended to accommodate lower-temperature regimes ($T < 10^4 \mathrm{\ K}$) as provided by \citet{Goodman_2012_ApJ}, following the methodology outlined in the literature \citep{Iijima_2016_PhD, Shoda_2021_AA}.

\vspace{1em}
\subsection{Phenomenology of flux emergence and interchange reconnection \label{sec:phenomenology_flux_emergence}}

In Eq.~\eqref{eq:source_terms}, the term $Q^{\mathrm{FE}}$ denotes the plasma heating rate attributed to interchange reconnection.
\rev{In this study, we consider interchange reconnection as a source of additional heating and disregard changes in flux tube shape and coronal-base parameters due to altered field connectivity.}
The energy released by this process is presumed to be conveyed through flux emergence, thereby linking $Q^{\mathrm{FE}}$ with the energy flux of flux emergence, represented as $F^{\mathrm{FE}}$. 
The relationship between these variables can be expressed as follows:
\begin{align}
    Q^{\mathrm{FE}} = - \frac{1}{r^2 f^{\mathrm{open}}} \frac{\partial}{\partial r} \left( F^{\mathrm{FE}} r^2 f^{\mathrm{open}} \right).
    \label{eq:heating_rate_flux_emergence}
\end{align}

Considering that flux emergence and interchange reconnection are inherently multi-dimensional processes, we offer a phenomenological depiction of $F^{\rm FE}$, subject to the ensuing assumptions. 

Primarily, we hypothesize that $F^{\rm FE}$ remains temporally invariant. 
Although the time-dependent nature of flux emergence frequently engenders various transient phenomena, our primary interest resides in the time-averaged properties of the stellar wind. 
As a result, we exclusively investigate the time-averaged energy transport mediated by flux emergence. 

Secondly, we contend that the energy flux associated with flux emergence bears a direct proportionality to the coronal background magnetic field, as substantiated by the analysis of the coronal magnetic field's recycling timescale \citep{Wang_2020_ApJ, Wang_2022_SolPhys}.
Specifically, we propose that the energy flux discerned at the coronal base is represented by
\begin{align}
    F^{\rm FE}_{\rm cb} = 1.0 \times 10^5 \left( \frac{B_{r,{\rm cb}}}{10 {\rm \ G}} \right) {\rm \ erg \ cm^{-2} \ s^{-1}}, \label{eq:energy_flux_emerging_flux_cb}
\end{align}
where $X_{\rm cb}$ represents the value of $X$ measured at the base of the corona, specifically defined as the location where the time-averaged temperature reaches $4.0 \times 10^5{\rm K}$ (see also Section~\ref{sec:dependence_on_Phi_surf}).

We remark that the $F^{\rm FE}_{\rm cb}$ value for $B_{r,{\rm cb}} = 10 {\rm \ G}$ is one-third of the initial value presented by \citet{Wang_2020_ApJ} ($2.9 \times 10^5 {\rm \ erg \ cm^{-2} \ s^{-1}}$). 
This deviation arises due to different estimates of the loop height $h_{\rm ER}$, typically associated with the total magnetic flux $\Phi_{\rm ER}$ as follows \citep{Cranmer_2010_ApJ_can_the_solar_wind}:
\begin{align}
    h_{\rm ER} = 2.0 \times 10^3 \left( \frac{\Phi_{\rm ER}}{1 \times 10^{18} {\rm \ Mx}} \right) {\rm \ km}
\end{align}
In \citet{Wang_2020_ApJ}, the mean loop height $\langle h_{\rm ER} \rangle = 6.4 \times 10^3 {\rm \ km}$ corresponds to the average magnetic flux $\langle \Phi_{\rm ER} \rangle = 1.0 \times 10^{19} {\rm \ Mx}$. 
Given the higher prevalence of flux emergence with smaller magnetic flux \citep{Thornton_2011_SolPhys}, the mean magnetic flux may decrease, implying an average loop height less than $6.4 \times 10^3 {\rm \ km}$ \citep{Wiegelmann_2004_SolPhys, Cranmer_2010_ApJ_can_the_solar_wind}.
In addition, considering magnetic loops lower than the transition region do not directly contribute to coronal heating, emerging flux with total magnetic flux less than $1.0 \times 10^{18} {\rm \ Mx}$ should be disregarded. 
Based on these premises, we estimate $\langle h_{\rm ER} \rangle = 2.0 \times 10^3 {\rm \ km}$. 
As the energy flux of interchange reconnection is assumed to vary with $\langle h_{\rm ER} \rangle$ \citep{Wang_2020_ApJ}, we lower the $F^{\rm FE}_{\rm cb}$ value by a factor of $6.4/2.0 \approx 3$.

Considering the aforementioned premises, the functional form of $F_{\rm FE}$ is delineated as follows:
\begin{align}
    \frac{F^{\rm FE}}{F^{\rm FE}_{\rm cb}} = \frac{r_{\rm cb}^2 f^{\rm open}_{\rm cb}}{r^2 f^{\rm open}} \frac{\exp \left( \frac{r^{\rm FE}-r}{\sigma^{\rm FE}} \right)}{1 + \exp \left( \frac{r^{\rm FE}-r}{\sigma^{\rm FE}} \right)}, \label{eq:energy_flux_emerging_flux}
\end{align}
where we adopt the values $r^{\rm FE}/r_\ast = 1.2$ and $\sigma^{\rm FE}/r_\ast = 0.02$ to ensure that the heating due to interchange reconnection is confined to the low (subsonic) corona.
\rev{We have investigated the dependence of $\dot{M}_w$ on $r^{\rm FE}$ and $\sigma^{\rm FE}$, finding it largely unchanged as long as heating by flux emergence is confined to the subsonic region (see Appendix~\ref{app:dependence_on_FE_parameters} for detail).}

\begin{figure}[t!]
\centering
\includegraphics[width=75mm]{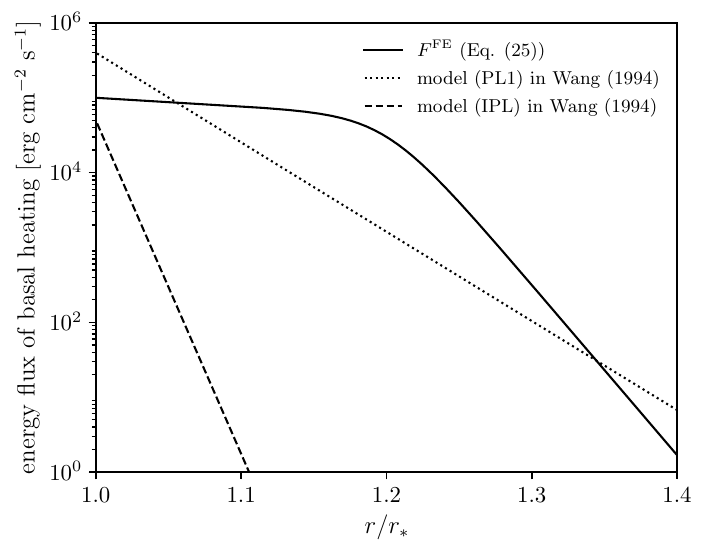}
\caption{
Comparison between $F^{\rm FE}$ (solid line, as given in Eq.~\eqref{eq:energy_flux_emerging_flux}) and the basal-heating energy fluxes utilized in \citet{Wang_1994_ApJ}. 
Dotted and dashed lines signify the plume (PL1) and interplume (IPL) models in \citet{Wang_1994_ApJ}, respectively.
}
\label{fig:FE_energy_flux_comparison}
\vspace{0em}
\end{figure}

It is noteworthy that a previous study \citep{Wang_1994_ApJ} incorporated an energy flux similar to Eq.~\eqref{eq:energy_flux_emerging_flux}.
A comparison between our flux-emergence energy flux (solid line, given by Eq.\eqref{eq:energy_flux_emerging_flux}) and the basal-heating energy flux from \citet{Wang_1994_ApJ} for both interplume (IPL, dashed line) and plume (PL1, dotted line) cases is shown in Figure~\ref{fig:FE_energy_flux_comparison}
The energy flux in our model finds its position between the IPL and PL1 models, with respect to the basal value and the damping length.
Given the successful modeling of plume and interplume in \citet{Wang_1994_ApJ}, it is plausible that our basal heating likely mirrors the actual one.

\vspace{1em}
\subsection{Boundary condition \label{sec:boundary_condition}}

The simulation domain encompasses the region from the stellar surface, constituting the inner boundary, to beyond the super-Alfv\'enic region, which serves as the outer boundary. 
We impose the free boundary condition at the outer boundary, as all MHD waves exhibit outward propagation. 
For the inner boundary, we utilize the fixed boundary condition for $\rho$, $p$, and $B_r$, as follows:
\begin{gather}
\rho_{\rm in} = 1.9 \times 10^{-7} {\rm \ g \ cm^{-3}}, \notag \\
p_{\rm in} = 9.4 \times 10^4 {\rm \ dyne \ cm^{-2}}, \notag \\
B_{r,{\rm in}} = 1.3 \times 10^3 {\rm \ G}, \notag
\end{gather}
where $X_{\rm in}$ represents the value of $X$ at the inner boundary. 
Additionally, we enforce a zero vertical velocity at the inner boundary: $v_{r,{\rm in}}=0$.

To impose the horizontal velocity and magnetic field such that 1. the reflected (inward) Alfv\'en waves penetrate the lower boundary, and 2. a (nearly) constant net Poynting flux is injected from the surface, we apply a free boundary condition for the inward Els\"asser variables:
\begin{align}
    \left. \frac{\partial}{\partial r} \boldsymbol{z}_\perp^{-} \right|_{\rm in} = 0,
\end{align}
where $\boldsymbol{z}_\perp^\pm = \boldsymbol{v}_\perp \mp \boldsymbol{B}_\perp/\sqrt{4 \pi \rho}$ are the Els\"asser variables. 
Subsequently, we adjust the amplitude of $\boldsymbol{z}_{\perp,{\rm in}}^+$ to maintain a fixed time-averaged wave energy flux \citep{Shoda_2020_ApJ}, the value of which is taken from the literature \citep{Cranmer_2005_ApJ}:
\begin{align}
    F^{\rm AW}_{\rm in} &= 5.0 \times 10^8 {\rm \ erg \ cm^{-2} \ s^{-1}}, \label{eq:Alfven_wave_energy_flux_photosphere}
\end{align}
where 
\begin{align}
    F^{\rm AW} = \overline{\left( \frac{1}{2} \rho \boldsymbol{v}_\perp^2 + \frac{\boldsymbol{B}_\perp^2}{4 \pi} \right) v_r - \frac{B_r}{4 \pi} \left( \boldsymbol{v}_\perp \cdot \boldsymbol{B}_\perp \right)}, \label{eq:energy_flux_Alfven_wave}
\end{align}
and $\overline{X}$ represents the time-average of $X$.

The spectra for $\boldsymbol{z}^+_{\perp,{\rm in}}$ are provided as follows:
\begin{gather}
    z_{x,y,{\rm in}}^+ \propto \sum_{i=0}^N \frac{\sin \left( 2 \pi f_i t + \phi_{i,x,y} \right)}{{f_i}^{5/6}}, \\
    f_i = \frac{\left( N-i\right) f_{\rm min} + i f_{\rm max}}{N},
\end{gather}
where $\phi_{i,x,y}$ is a random phase function, and $N+1$ designates the total mode count. Furthermore,
\begin{align}
    f_{\rm min} = 1.00 \times 10^{-3} {\rm \ Hz}, \ \ \ f_{\rm max} = 1.00 \times 10^{-2} {\rm \ Hz},
\end{align}
define the least and greatest wave frequencies. In our study, $N$ is set to 20.
This implementation results in the outward Els\"asser energy exhibiting a Kolmogorov-type spectrum ($ {\boldsymbol{z}^+_{\perp,{\rm in}} }^2 \propto f^{-5/3}$), as observed in the spectrum of the coronal transverse waves \citep{Morton_2019_Nature_Astronomy}.

\subsection{Simulation setup and numerical solver}

The free parameters in this study, the total unsigned magnetic flux on the surface, $\Phi^{\rm surf}$, and the open-to-surface magnetic flux ratio ($\xi^{\rm open} = \Phi^{\rm open}/\Phi^{\rm surf}$), dictate the shape of the background flux tube as per Eq.s~\eqref{eq:fopen}--\eqref{eq:br_radial_profile}. 
When representing dependencies on $\Phi^{\rm surf}$, it is normalized to the solar value, specifically $\Phi^{\rm surf}_\odot = 3.0 \times 10^{23}$ Mx.
We execute a total of 24 simulations, comprised of six variations for $\Phi^{\rm surf} / \Phi^{\rm surf}_\odot$ ($\Phi^{\rm surf} / \Phi^{\rm surf}_\odot = 1.36, \ 2.72, \ 5.43, \ 10.86, \ 21.72, \ 43.44$) and four for $\xi^{\rm open}$ ($\xi^{\rm open} = 0.05, \ 0.1, \ 0.2, \ 0.4$).

The outer boundary of the simulation domain, denoted $r_{\rm out}$, is adjusted to situate in the super Alfv\'enic region (in a quasi-steady state). 
Specifically, $r_{\rm out} / r_\ast = 44.9$ for $\Phi^{\rm surf} / \Phi^{\rm surf}_\odot = 1.36$ and $r_{\rm out} / r_\ast = 201-270$ (depending on $\xi^{\rm open}$) for $\Phi^{\rm surf} / \Phi^{\rm surf}_\odot = 43.44$. 
The simulation time, represented as $\tau_{\rm sim}$, is dependent on the domain size due to the increased duration required for larger domains to reach a quasi-steady state. 
Hence, we set $\tau_{\rm sim} = 4.8 \times 10^5 {\rm \ s}$ for $\Phi^{\rm surf} / \Phi^{\rm surf}_\odot = 1.36$ and $\tau_{\rm sim} = 7.2 \times 10^5 {\rm \ s}$ for $\Phi^{\rm surf} / \Phi^{\rm surf}_\odot = 43.44$. 
Regardless of $\tau_{\rm sim}$, time-averaged values are computed over a constant period $\tau_{\rm ave} = 1.8 \times 10^5 {\rm \ s} = 2.08 {\rm \ day}$.

The iterative computation of grid spacing is carried out as delineated below:
\begin{gather}
    \frac{\Delta r^i}{\Delta r_{\rm min}} = \max \left[1, \ \min \left [ \frac{\Delta r_{\rm max}}{\Delta r_{\rm min}}, \ \frac{2\varepsilon}{2 + \varepsilon} \frac{r^{i-1}-r_{\rm exp}}{\Delta r_{\rm min}}+1 \right] \right], \notag \\
    r^i = r^{i-1} + \frac{1}{2} \left( \Delta r^{i-1} + \Delta r^i \right),
\end{gather}
where $r^i$ and $\Delta r^i$ represent the radial distance of the $i$-th grid center and the $i$-th grid size, respectively.
The parameter $r_{\rm exp}$ signifies the radial distance beyond which grid expansion occurs, and $\varepsilon$ denotes the ratio of the neighboring grid size in the grid-expansion region.
$\Delta r_{\rm min}$ is held constant at $20 {\rm \ km}$, whereas $\Delta r_{\rm max}$ varies within the range of $2000-6000 {\rm \ km}$ contingent upon the value of $\Phi^{\rm surf}$.
In this work, we establish $\Delta r^0 = \Delta r_{\rm min}$, $r^0 = r_\ast$, $r^{\rm exp}/r_\ast = 1.0144$, and $\varepsilon = 0.0158$.

In numerically solving Eqs.~\eqref{eq:basic_conservation_law}--\eqref{eq:source_terms}, we employ the HLLD solver \citep{Miyoshi_2005_JCP} in conjunction with the third-order SSP Runge--Kutta integration \citep{Shu_1988_JCP, Gottlieb_2001_SIAMR}.
We adopt the cross-section-weighted variables \citep{Shoda_2021_AA} to render the form of the fundamental equations analogous to the conventional one-dimensional MHD equation. 
To guarantee elevated spatial accuracy within $r<r^{\rm exp}$, the fifth-order MP5 reconstruction \citep{Suresh_1997_JCP} is implemented, whereas the second-order MUSCL reconstruction \citep{van_Leer_1979_JCP} is utilized in the grid-expansion region. 
In the area where $\Delta r = \Delta r_{\rm max}$, the MP5 reconstruction is applied to transverse velocity and magnetic field, while the MUSCL reconstruction is employed for the remaining variables. Thermal conduction is solved separately using the super-timestepping method \citep{Meyer_2012_MNRAS, Meyer_2014_JCP}.

\section{Result \label{sec:result}}

\subsection{Dependence on the surface magnetic flux \label{sec:dependence_on_Phi_surf}}

\begin{figure}[t!]
\centering
\includegraphics[width=75mm]{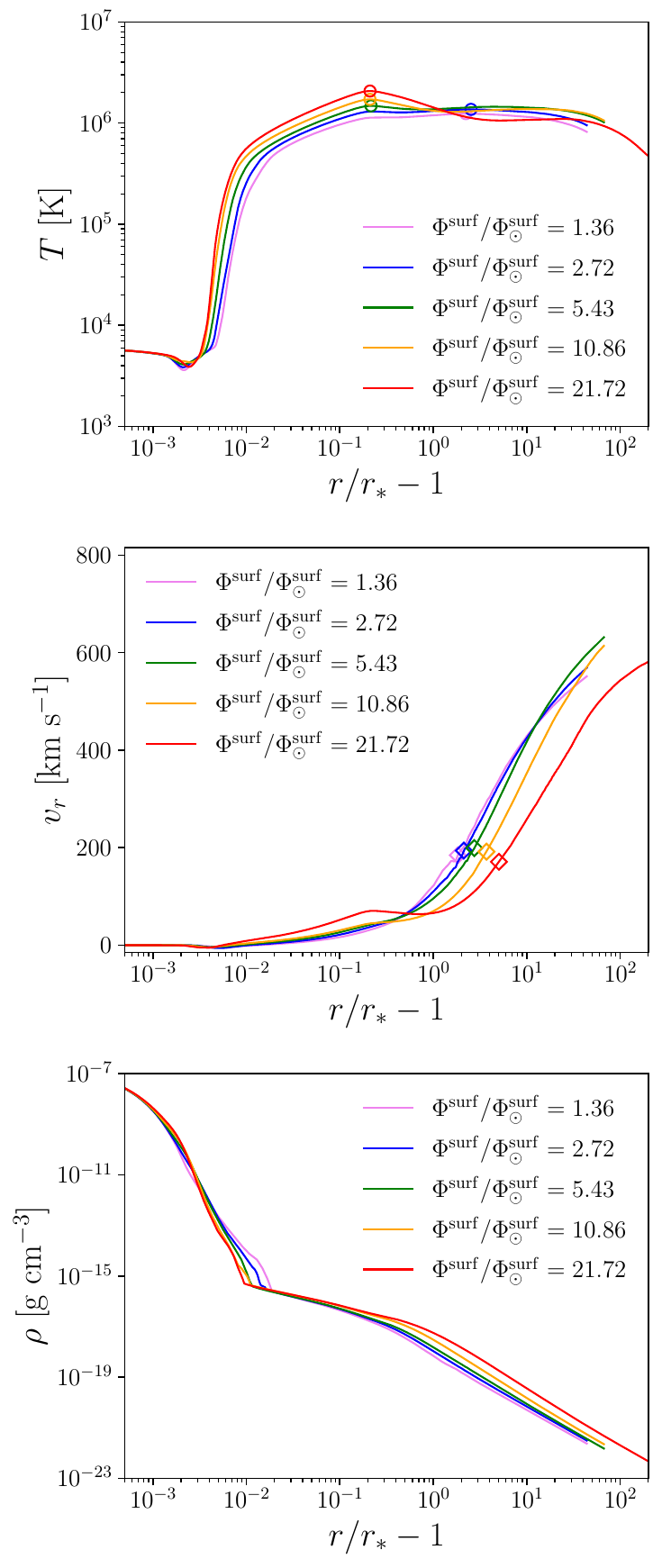}
\caption{
Time-averaged radial profiles of stellar-wind parameters: mass density (top panel), temperature (middle panel), and radial velocity (bottom panel). 
Line color signifies the corresponding surface magnetic flux. 
Temperature-maximum points are denoted by circles in the top panel, while sonic points are represented by diamonds in the middle panel. 
The open-to-surface magnetic flux ratio is maintained at a constant value of $\xi^{\rm open} = 0.2$.
}
\label{fig:overview_Phi_dependence}
\vspace{0em}
\end{figure}

Figure \ref{fig:overview_Phi_dependence} presents a comparison of simulation outcomes for varying surface magnetic flux magnitudes.
A constant open-to-surface magnetic flux ratio is maintained ($\xi^{\rm open} = 0.2$).
Each subplot illustrates the time-averaged radial profiles for temperature (upper), velocity (central), and density (lower).
Lines exhibiting a more intense red hue signify a greater surface magnetic flux.
Circles situated in the upper subplot represent the point of maximum temperature.
The position of the sonic point (defined as $v_r = \sqrt{\gamma p/\rho}$) is indicated by diamonds in the central subplot.

The influence of flux emergence and interchange reconnection on temperature profiles is distinctly evident.
For $\Phi^{\rm surf}/\Phi^{\rm surf}_\odot \lesssim 4$, the temperature attains its maximum value at $r/r_\ast \approx 3.5$, where the solar wind undergoes efficient acceleration.
Conversely, when $\Phi^{\rm surf}/\Phi^{\rm surf}_\odot \gtrsim 4$, the temperature reaches its peak at $r/r_\ast \approx 1.2$, the location at which the majority of the emerging-flux energy is transformed into heat.
This increased heat deposition due to flux emergence could potentially result in a two-step acceleration process for the stellar wind, as observed in the case of $\Phi^{\rm surf}/\Phi^{\rm surf}_\odot = 21.72$.
This acceleration occurs first in close proximity to the Sun, at $r/r_\ast \approx 1.1$, and is driven by the gas-pressure force \citep{Parker_1958_ApJ, Kopp_1976_SolPhys}, followed by a second acceleration in the supersonic region, at $r/r_\ast \approx 10$, propelled by the wave-pressure force \citep{Alazraki_1971_AA, Belcher_1971_ApJ}.
It is important to note that this initial phase of wind acceleration may manifest as the active-region outflow observed within the solar corona \citep{Sakao_2007_Science, Harra_2008_ApJ, Brooks_2015_NatreCommunications}.
We find a comparable non-monotonic trend in $v_r$ as a function of $r$ in the previous models \citep{Cranmer_2007_ApJ, Cranmer_2010_ApJ_an_efficient_approximation, Panasenco_2019_ApJ}, as identified in the case where the flux-tube expansion varied non-monotonically with $r$.

In the profiles of $v_r$, it is evident that the stellar-wind velocity at the outer boundary remains relatively constant, 
despite the considerable diversity in open magnetic flux. 
Solar wind observations suggest that the governing factor for wind velocity is the flux-tube expansion factor \citep{Wang_1990_ApJ, Arge_2000_JGR, Riley_2015_SpaceWeather}, rather than the open magnetic flux \citep{Wang_2020_ApJ}. 
In the proposed model, this expansion factor is equivalent to $1/\xi^{\rm open}$ and remains fixed, as illustrated in Figure \ref{fig:overview_Phi_dependence}.
Consequently, our model offers a coherent explanation for the observed behavior of the solar wind.

The mass density of the solar wind exhibits a monotonic increase as a function of $\Phi^{\rm surf}$.
This can be attributed to the fact that the majority of emerging-flux heating transpires below the subsonic point, thereby leading to an enhancement in mass density at the critical point and further along \citep{Leer_1980_JGR, Pneuman_1980_AA, Hansteen_2012_SSRev}.
Consequently, the mass-loss rate, denoted by $\dot{M}_w$, displays a monotonic increase with $\Phi^{\rm surf}$, provided that $\xi^{\rm open} \equiv \Phi^{\rm open}/\Phi^{\rm surf}$ remains constant.
A comprehensive discussion on the behavior of $\dot{M}_w$ in relation to magnetic parameters can be found in Section~\ref{sec:scaling_law}.

\begin{figure}[t!]
\centering
\includegraphics[width=75mm]{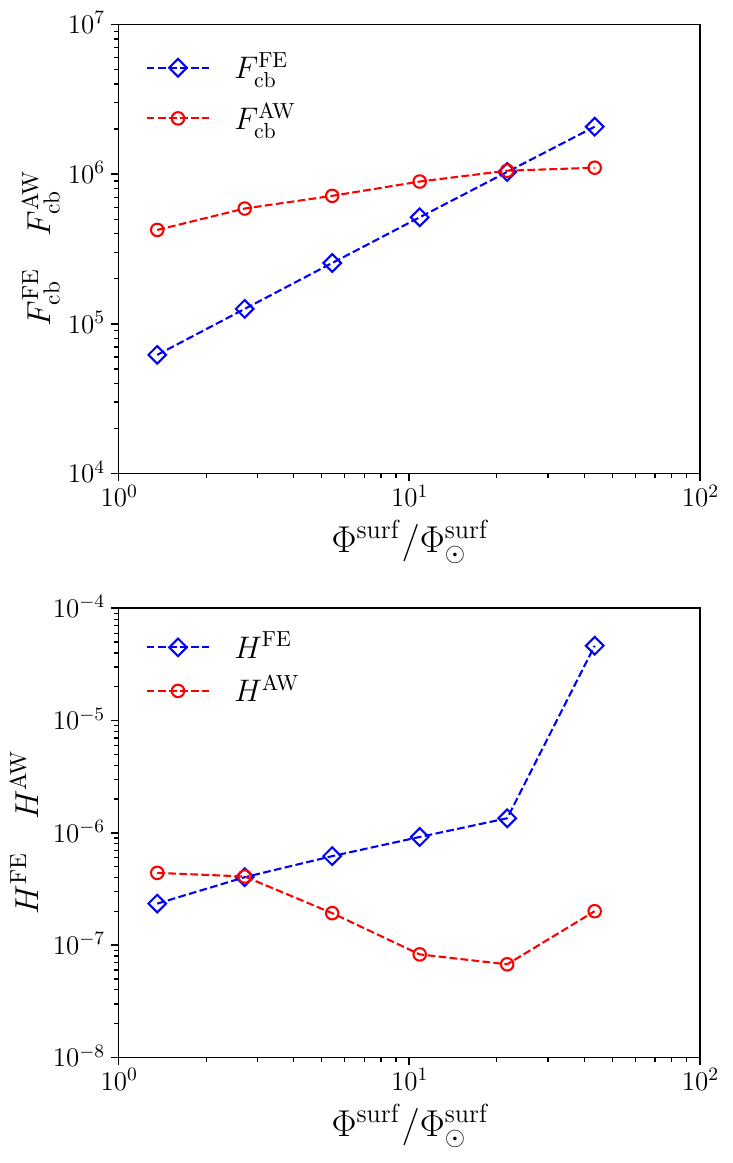}
\caption{
Upper panel: Coronal-base energy flux plotted against $\Phi^{\rm surf}/\Phi^{\rm surf}_\odot$. 
The blue diamonds denote the energy flux of the emerging flux ($F^{\rm FE}_{\rm cb}$), while the red circles indicate the Alfv\'en wave energy flux ($F^{\rm AW}_{\rm cb}$). 
Lower panel:  heating rate in the subsonic corona plotted against $\Phi^{\rm surf}/\Phi^{\rm surf}_\odot$. 
Blue diamonds and red circles represent heating rates resulting from the emerging flux ($H^{\rm FE}$) and Alfv\'en wave ($H^{\rm AW}$), respectively.
}
\label{fig:energy_flux_heating_comparison}
\vspace{1em}
\end{figure}

The substantial diversity seen in Figure~\ref{fig:overview_Phi_dependence} indicates a primary reliance of the wind-driving mechanisms on $\Phi^{\rm surf}$. 
We verify this by computing the energy fluxes due to flux emergence ($F^{\rm FE}$, see Eq.~\eqref{eq:energy_flux_emerging_flux}) and Alfv\'en waves ($F^{\rm AW}$, see Eq.~\eqref{eq:energy_flux_Alfven_wave}) at the coronal base, $r = r_{\rm cb}$, where the average temperature is 0.4 million Kelvin ($\overline{T} = 4.0 \times 10^5 {\rm \ K}$).
The top panel in Figure~\ref{fig:energy_flux_heating_comparison} plots $F^{\rm FE}_{\rm cb}$ (blue diamonds) and $F^{\rm AW}_{\rm cb}$ (red circles) against $\Phi^{\rm surf} / \Phi^{\rm surf}_\odot$. 
Flux-emergence energy flux proportionally rises with $\Phi^{\rm surf}$, as shown by Eq.~\eqref{eq:energy_flux_emerging_flux_cb}, while Alfv\'en-wave energy flux gently ascends and plateaus at high $\Phi^{\rm surf}$, consistent with \citet{Shoda_2020_ApJ}. 
Consequently, $F^{\rm FE}_{\rm cb}$ surpasses $F^{\rm AW}_{\rm cb}$ at $\Phi^{\rm surf} / \Phi^{\rm surf}_\odot \approx 20$.

This does not mean, however, that the influence of flux emergence is negligible in $\Phi^{\rm surf} / \Phi^{\rm surf}_\odot < 20$. 
Given that mass loss is affected by the heat deposited beneath the sonic critical point \citep{Leer_1980_JGR, Pneuman_1980_AA, Hansteen_2012_SSRev}, the heating rate in the subsonic corona is crucial.
We thus calculate the averaged heating rates from Alfv\'en waves ($H^{\rm AW}$) and flux emergence ($H^{\rm FE}$) in the subsonic corona as follows:
\begin{gather}
    H^{\rm AW} = - \frac{1}{r_{\rm sc} - r_{\rm cb}} \int_{r_{\rm cb}}^{r_{\rm sc}} dr \ \frac{1}{r^2 f^{\rm open}} \left( F^{\rm AW} r^2 f^{\rm open} \right), \label{eq:heating_rate_AW_subsonic_corona} \\[0.5em]
    H^{\rm FE} = - \frac{1}{r_{\rm sc} - r_{\rm cb}} \int_{r_{\rm cb}}^{r_{\rm sc}} dr \ \frac{1}{r^2 f^{\rm open}} \left( F^{\rm FE} r^2 f^{\rm open} \right), \label{eq:heating_rate_FE_subsonic_corona}
\end{gather}
where $r_{\rm sc}$ is the sonic critical point's radial distance. 
In the above calculation, we assume that the loss in the energy flux is attributed to heating.
\rev{The detailed radial profiles of heating and cooling rates are shown in Appendix~\ref{app:radial_profiles_of_heating_cooling}.}

The lower panel of Figure~\ref{fig:energy_flux_heating_comparison} plots $H^{\rm FE}$ (blue diamonds) and $H^{\rm AW}$ (red circles) against $\Phi^{\rm surf} / \Phi^{\rm surf}_\odot$. 
Unlike energy flux, the flux-emergence effect overtakes the Alfv\'en-wave effect at $\Phi^{\rm surf} / \Phi^{\rm surf}_\odot \approx 3$. 
This finding aligns with the observed regime transition in Figure \ref{fig:overview_Phi_dependence}. 
Thus, flux emergence could significantly impact the mass-loss rate, even when it plays a minor role in overall stellar-wind energetics.

\begin{figure}[t!]
\centering
\includegraphics[width=75mm]{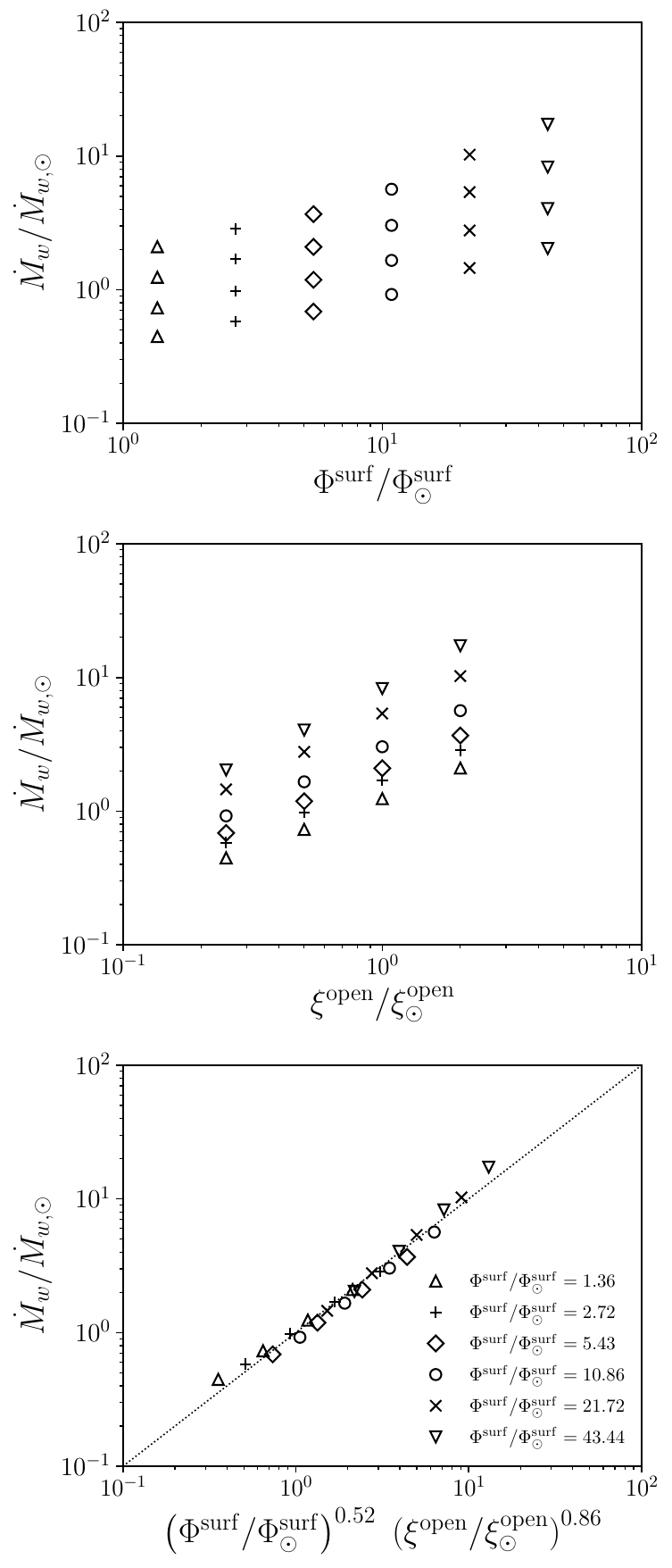}
\vspace{0.5em}
\caption{
Mass-loss rate ($\dot{M}_w$) as a function of magnetic-field parameters ($\Phi^{\rm surf}$ and $\xi^{\rm open}$) for the top and middle panels, respectively. 
Normalization parameters are set to solar values: $\dot{M}_{w,\odot}=2.0 \times 10^{-14} \ M_\odot \ {\rm yr}^{-1}$, $\Phi^{\rm surf}_\odot = 3.0 \times 10^{23} {\rm \ Mx}$, and $\xi^{\rm open}_\odot = 0.2$. 
The bottom panel displays the double power-law fit to the simulation data, as described by Eq.~\eqref{eq:scaling_relation}.
}
\label{fig:Mdot_compiled}
\end{figure}

\subsection{Theoretical scaling law of $\dot{M}_w$ \label{sec:scaling_law}}

Figure \ref{fig:Mdot_compiled} displays the simulated data for $\dot{M}_w$ across a range of $\Phi^{\rm surf}$ and $\xi^{\rm open}$. 
The three parameters are normalized using solar reference values: $\dot{M}_{w,\odot}=2.0 \times 10^{-14} \ M_\odot \ {\rm yr}^{-1}$, $\Phi^{\rm surf}_\odot = 3.0 \times 10^{23} {\rm \ Mx}$, and $\xi^{\rm open}_\odot = 0.2$. 
The top and middle panels indicate that $\dot{M}_w$ increases with $\Phi^{\rm surf}$ and $\xi^{\rm open}$ in a manner approximating a power-law relationship. 
Inspired by these trends, we fit the simulation data employing double power-law relations with respect to $\Phi^{\rm surf}$ and $\xi^{\rm open}$. Specifically, the fitting results yield the following scaling law:
\begin{align}
    \frac{\dot{M}_w}{\dot{M}_{w,\odot}} = \left( \frac{\Phi^{\rm surf}}{\Phi^{\rm surf}_\odot} \right)^{0.52} \left( \frac{\xi^{\rm open}}{\xi^{\rm open}_\odot} \right)^{0.86}. \label{eq:scaling_relation}
\end{align}
The bottom panel of Figure~\ref{fig:Mdot_compiled} illustrates the comparison between the scaling law in Eq.~\eqref{eq:scaling_relation} and the simulation data. 
The results demonstrate the efficacy of Eq.\eqref{eq:scaling_relation} in accurately representing the simulation data.

\begin{figure}[t!]
\centering
\includegraphics[width=75mm]{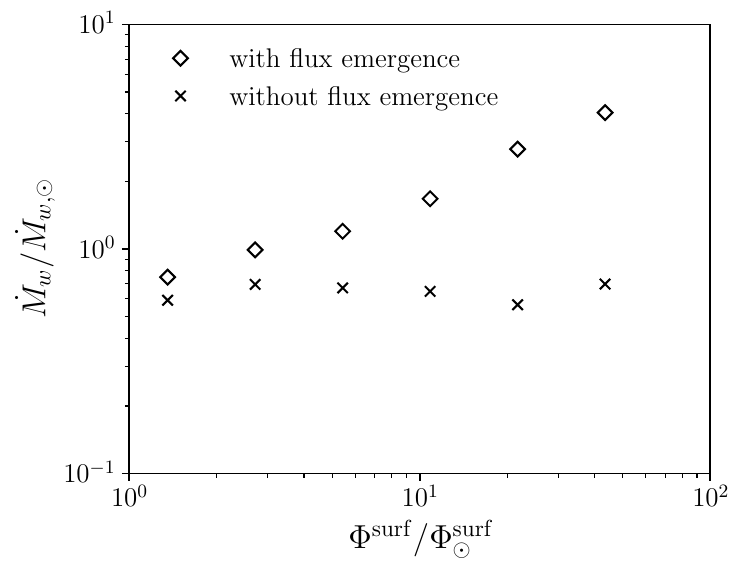}
\caption{
Comparative analysis of the $\Phi^{\rm surf}$--$\dot{M}_w$ relations incorporating and excluding the influence of flux emergence, maintaining a constant open-to-surface flux ratio at $\xi^{\rm open} = 0.1$.
}
\label{fig:comparison_Mdot_w_wo_FE}
\vspace{0em}
\end{figure}

The correlation between stellar surface magnetic flux, denoted as $\Phi^{\rm surf}$, and stellar X-ray luminosity, represented by $L_{\rm X}$, is typified by power-law relations \citep{Pevtsov_2003_ApJ,Kochukhov_2020_AA,Zhuleku_2020_AA}. 
As a result, the scaling law expressed in Eq.~\eqref{eq:scaling_relation} can be recast in terms of $L_{\rm X}$. 
For example, \citet{Toriumi_2022_ApJ} and \citet{Toriumi_2022_ApJS} illustrate that
\begin{align}
    \frac{L_{\rm X}}{L_{{\rm X},\odot}} = \left( \frac{\Phi^{\rm surf}}{\Phi^{\rm surf}_\odot} \right)^{1.15}. \label{eq:toriumi_airapetian_2022}
\end{align}
By integrating Eq.s~\eqref{eq:scaling_relation} and~\eqref{eq:toriumi_airapetian_2022}, the resulting equation is
\begin{align}
    \frac{\dot{M}_w}{\dot{M}_{w,\odot}} = \left( \frac{L_{\rm X}}{L_{{\rm X},\odot}} \right)^{0.45} \left( \frac{\xi^{\rm open}}{\xi^{\rm open}_\odot} \right)^{0.91},
\end{align}
which displays congruity with the observation-based scaling law given by \citep{Vidotto_2021_LRSP}:
\begin{align}
    \frac{\dot{M}_w}{\dot{M}_{w,\odot}} = \left( \frac{L_{\rm X}}{L_{{\rm X},\odot}} \right)^{0.66}, \label{eq:Lx_Mdot_empirical}
\end{align}
provided that $\xi^{\rm open}$ does not diminish substantially with $L_{\rm X}$.

In order to elucidate the impact of flux emergence on the scaling law of the mass loss rate ($\dot{M}_w$), 
we present in Figure~\ref{fig:comparison_Mdot_w_wo_FE} a comparison of the dependence of $\dot{M}_w$ on the surface magnetic flux ($\Phi^{\rm surf}$) for scenarios incorporating flux emergence and those excluding it ($F^{\rm FE} = 0$). 
For the purpose of this analysis, we maintain a constant open-to-surface flux ratio of $\xi^{\rm open} = 0.1$. 
In the absence of flux emergence, it is observed that $\dot{M}_w$ exhibits a nearly constant relationship with $\Phi^{\rm surf}$, in stark contrast to the monotonically increasing trend observed when flux emergence is present. It is noteworthy that the near-constant behavior of $\dot{M}_w$ aligns with the results of previous stellar-wind simulations \citep{Shoda_2020_ApJ}. 
Thus, the observed enhancement in $\dot{M}_w$ corresponding to an increase in activity level \citep{Wood_2021_ApJ,Vidotto_2021_LRSP} be ascribed to the basal heating engendered by flux emergence.

\subsection{Comparison with the solar wind observations}

\begin{figure}[t!]
\centering
\includegraphics[width=75mm]{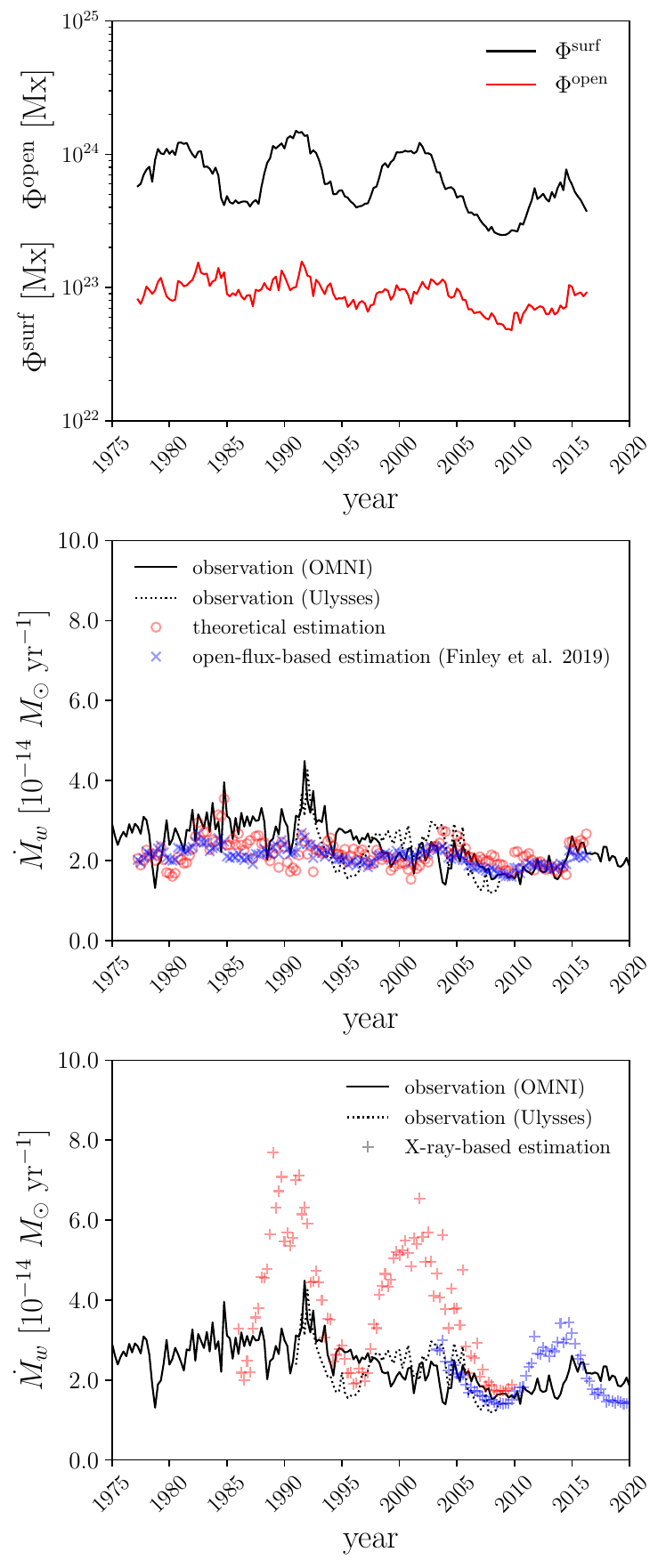}
\caption{
Long-term trends (quarterly averages) of solar magnetic field and solar wind parameters. 
Upper panel: Surface ($\Phi^{\rm surf}$, displayed in black) and open ($\Phi^{\rm open}$, in red) solar magnetic flux. 
Central panel: The solar wind mass-loss rate inferred from OMNI (solid line) and Ulysses (dotted line) datasets, juxtaposed with \rev{the open-flux-based fitting (blue crosses, refer to the text for detail)} and theoretical mass-loss rate (red circles). 
Lower panel: Similar to the central panel, however, observations are counterpointed with an X-ray-based empirical scaling law (denoted by pluses), where different colors specify different X-ray data sources (refer to the text for detail).
}
\label{fig:comparison_with_observation}
\vspace{0em}
\end{figure}

To substantiate the theoretical scaling law presented in Eq.~\eqref{eq:scaling_relation}, we examine its compatibility with solar observations. Our objective is to juxtapose the long-term trends of mass-loss rates ascertained through Eq.~\eqref{eq:scaling_relation} and those directly measured in the solar wind.

Solar-disk observations spanning several decades provide data on the total unsigned magnetic flux of the Sun. 
While the open magnetic flux is not directly observable, it can be inferred either through magnetic-field extrapolation techniques \citep{Wang_2002_JGR} or via in-situ measurements in the solar wind, presuming that the $r^2 B_r$ is nearly uniform in latitude \citep{Smith_1995_GRL}. 
The discrepancy between these two estimation methods is currently acknowledged as the open-flux problem \citep{Linker_2017_ApJ, Linker_2021_ApJ, Arge_2023_arXiv}. 
In this study, we procure the observed data for $\Phi^{\rm surf}$ and $\Phi^{\rm open}$ from the literature \citep{Cranmer_2017_ApJ}, which have been averaged over 0.25-year intervals.

We estimate the mass-loss rate from in-situ measurements in a manner analogous to the open flux determination, i.e., by assuming that $r^2 \rho v_r$ remains constant in latitude. 
However, it is important to note that the scaled momentum flux ($r^2 \rho v_r^2$) exhibits a uniform distribution rather than the scaled mass flux ($r^2 \rho v_r$) \citep{Goldstein_1995_SSRev, Goldstein_1996_AA, Le_Chat_2012_SolPhys}. 
Owing to the reduced wind speed within the ecliptic plane, computing $\dot{M}_w$ from the 1-au (OMNI) data could potentially result in overestimation. 
To ascertain whether a significant systematic error arises due to the sampling bias in latitude, we calculate the mass-loss rate utilizing both OMNI and Ulysses data.

In addition to the theoretical scaling law in Eq.~\eqref{eq:scaling_relation}, we also evaluate the compatibility of the stellar-based empirical scaling law in Eq.~\eqref{eq:Lx_Mdot_empirical} with solar observations. 
Since $L_{\rm X}$ in Eq.~\eqref{eq:Lx_Mdot_empirical} represents the X-ray luminosity measured in the ROSAT band \citep[$5.2-124 {\rm \ \AA}$,][]{Truemper_1982_AdSpR,Gudel_2007_LRSP}, we employ the daily spectral irradiance data from the XUV Photometer System \citep[XPS,][]{Woods_2005_SolPhys_XPS_overview_calibration, Woods_2005_SolPhys_XPS_SORCE} onboard the Solar Radiation and Climate Experiment (SORCE) satellite to compute the solar $L_{\rm X}$.
In addition, we also estimate the solar $L_{\rm X}$ using an empirical conversion from the GOES band ($1-8 {\rm \ \AA}$) the ROSAT band \citep{Judge_2003_ApJ, Cranmer_2021_AAS}.
\rev{The solar X-ray luminosity is set to $L_{{\rm X},\odot}=1.0 \times 10^{27} {\rm \ erg \ s^{-1}}$ \citep{Peres_2000_ApJ, Johnstone_2015_AA}.}

Figure~\ref{fig:comparison_with_observation} illustrates the time evolution of three-month-averaged parameters pertinent to the solar magnetic field and solar wind. The upper panel displays the filling factors of both the unsigned surface magnetic flux and open magnetic flux, which is congruent with Figure 2a of \citet{Cranmer_2017_ApJ}. 
\rev{
In the central panel, observed mass-loss rates are represented by lines, while estimated rates are denoted by symbols.
The solid and dashed lines represent the observed mass-loss rate as determined from 1-au (OMNI) and Ulysses data, respectively, exhibiting analogous behavior.
The red circles indicate theoretical rates derived from Eq.~\eqref{eq:scaling_relation}, and blue crosses represent the empirical mass-loss rate derived from data over five solar cycles \citep{Finley_2019_ApJ_solar_angular_momentum_loss_past_millenia}:
\begin{align}
    \dot{M}_w = 1.26 \times 10^{12} {\rm \ g \ s^{-1}} \left( \frac{\Phi^{\rm open}}{8.0 \times 10^{22} {\rm \ Mx}} \right)^{0.44}.\label{eq:Mdot_finley2019}
\end{align}
}
The lower panel is similar to the central panel, juxtaposing observations with the empirical X-ray-based estimations (indicated by pluses) from Eq.~\eqref{eq:Lx_Mdot_empirical}. 
Here, red and blue pluses correspond to $L_{\rm X}$ data ascertained from GOES and SORCE/XPS, respectively.

\rev{Our theoretical prediction better represents the long-term trend of $\dot{M}_w$ than the X-ray empirical law. 
Furthermore, it closely matches the empirical equation presented in Eq.~\eqref{eq:Mdot_finley2019}. 
Both our model and fitted relation demonstrate an enhanced agreement with observational data in the later phase (year $\ge 2000$), which may be attributed to the improved precision of magnetic-field measurements (refer to \citet{Cranmer_2017_ApJ} for an in-depth explanation). 
Notably, neither approach captures the observed $\dot{M}_w$ peak around 1991-1992, suggesting an anomalous phase of the solar wind during this period.
}

Based on this comparative analysis, we deduce that the hybrid model utilized in this study offers a suitable representation of solar (and potentially stellar) wind properties. 
Furthermore, we assert that Eq.~\eqref{eq:scaling_relation} provides a valuable tool for estimating the mass-loss rate of solar-type stars.
\rev{Investigating the applicability of our scaling law in the stellar context is beyond the scope of this study. 
However, it is pertinent to assess if our law offers any advantages over the X-ray-based scaling law or if both are analogous.}

\section{Discussion \label{sec:summary_discussion}}

In the present investigation, we introduce an innovative stellar-wind model founded upon the hybrid scenario, which incorporates both Alfv\'en waves and interchange reconnection processes. 
A comprehensive parameter survey leads to the derivation of a scaling relationship for $\dot{M}_w$, as presented in Eq.~\eqref{eq:scaling_relation}. 
This relationship aligns with the observed increase in $\dot{M}_w$ as a function of X-ray flux in low-mass stars and the near-constant nature of solar-wind mass flux, thereby offering potential utility for estimating mass-loss rates in solar-type stars.

The application of Eq.~\eqref{eq:scaling_relation} is complicated by the necessity of calculating the open magnetic flux, a task currently unattainable even for our Sun \citep{Linker_2017_ApJ}.
However, the latest improvements in modeling stellar large-scale magnetic field structures, via global MHD simulations and Zeeman-Doppler Imaging \citep{Folsom_2016_MNRAS, Folsom_2018_MNRAS, See_2019_ApJ}, indicate potential feasibility for deducing open magnetic flux \citep{OFionnagain_2019_MNRAS, Airapetian_2021_ApJ, Evensberget_2021_MNRAS, Evensberget_2022_MNRAS, Evensberget_2023_MNRAS}.
A comprehensive study of the relation between the Sun's surface and open magnetic flux \citep{Yoshida_2023_ApJ} could also be crucial in eliminating $\Phi^{\rm open}$ from Eq.~\eqref{eq:scaling_relation}.

While our model demonstrates efficacy, it is essential to acknowledge the presence of numerous free parameters that could potentially influence our conclusions.
A subset of these parameters is linked to the formulation of $F^{\rm FE}_{\rm cb}$, which represents the energy flux discharged by interchange reconnection in the corona, as delineated in Section~\ref{sec:phenomenology_flux_emergence}. 
Additionally, there exists an uncertainty in the net energy flux of Alfv\'en waves, as denoted by Eq.~\eqref{eq:Alfven_wave_energy_flux_photosphere}. 
Given the centrality of these parameters to our model, it is imperative that their values are validated or adjusted through three-dimensional simulations in future research.

Despite the aforementioned inherent limitations, it can be reasonably postulated that the influence of flux emergence may play a significant role in modulating mass-loss rates.
Consequently, it is meritorious to reexamine extant models concerning wind mass-loss rates predicated on the Alfv\'en-wave framework \citep{Cranmer_2011_ApJ, Shoda_2020_ApJ}.
Future stellar wind models should concurrently integrate effects of Alfv\'en wave and flux emergence, alongside stellar rotation \citep{Weber_1967_ApJ, Belcher_1976_ApJ, Holzwarth_2007_AA, Shoda_2020_ApJ}.

\rev{
Finally, we address the implications of our work for the magnetic braking problem in low-mass stars. 
Low-mass stars with Rossby numbers near the solar value are found to halt their spin-down \citep{van_Saders_2016_Nature, van_Saders_2019_ApJ, Metcalfe_2023_ApJ}, a phenomenon under active investigation in stellar physics. 
A plausible explanation is a shift in magnetic morphology, where higher-order multipolar components become dominant \citep{Garraffo_2016_AA, Garraffo_2018_ApJ}. 
This leads to reduced open magnetic flux \citep{Yoshida_2023_ApJ} and mass-loss rate as per Eq.~\eqref{eq:scaling_relation}, subsequently lowering magnetic braking efficiency \citep{Matt_2012_ApJ, Reville_2015_ApJ, Finley_2018_ApJ}.
Our findings therefore support this hypothesis, although we need further investigations such as deriving a formula for open magnetic flux as a function of surface magnetic field and mass-loss rate.
}

Numerical computations were carried out on the Cray XC50 at the Center for Computational Astrophysics (CfCA), National Astronomical Observatory of Japan.
This work made use of matplotlib, a Python library for publication quality graphics \citep{Hunter_2007_CSE}, and NumPy \citep{van_der_Walt_2011_CSE}.
This work was supported by JSPS KAKENHI Grant Nos. JP20KK0072 (PI: S. Toriumi), JP21H01124 (PI: T. Yokoyama), JP21H04492 (PI: K. Kusano), and JP22K14077 (PI: M. Shoda).

\begin{appendix}

\section{Dependence on the free parameters associated with flux emergence \label{app:dependence_on_FE_parameters}}

\begin{figure}[t!]
\centering
\includegraphics[width=75mm]{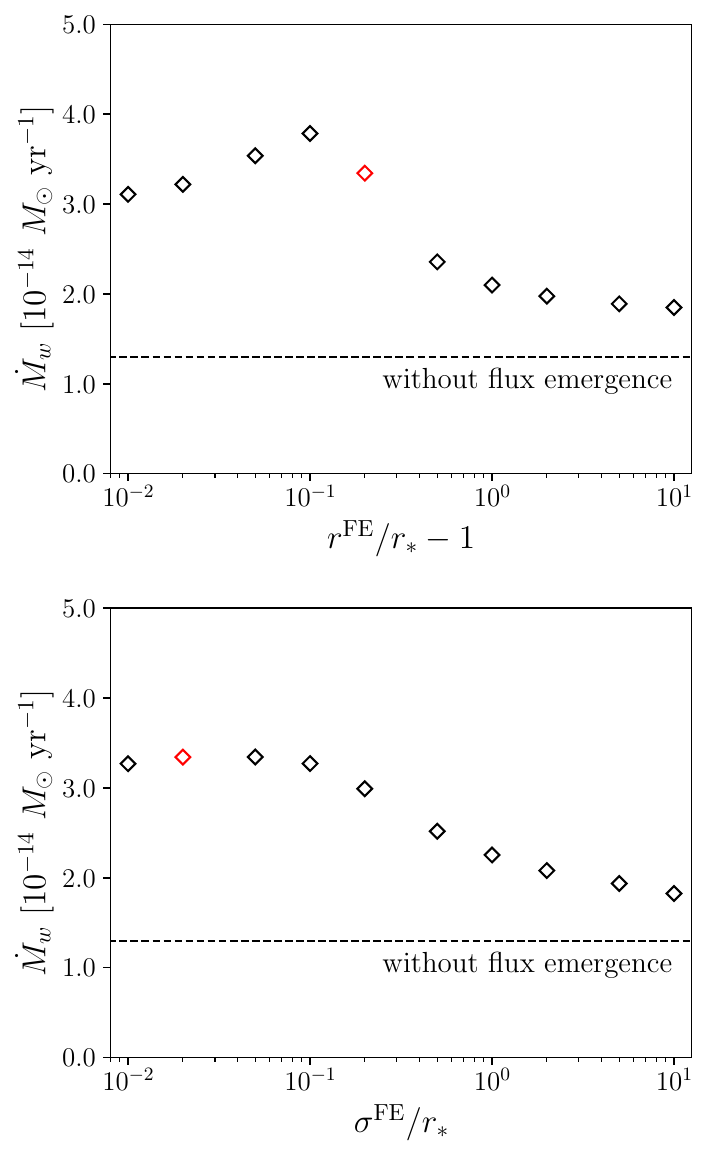}
\caption{
\rev{Dependence of mass-loss rate on flux-emergence heating parameters.
Top: Variation with radial distance of heat deposition ($r^{\rm FE}$).
Bottom: Variation with spatial extent of heat deposition ($\sigma^{\rm FE}$).
Reference cases are indicated by red symbols.
The surface magnetic flux and open-to-surface magnetic flux ratio are set at $\Phi^{\rm surf}/\Phi^{\rm surf}_\odot = 10.86$ and $\xi^{\rm open} = 0.1$, respectively.}
}
\label{fig:Mdot_dependence_on_rloop}
\end{figure}

\rev{
In this study, we augment the conventional Alfv\'en-wave solar wind model by incorporating the effects of flux emergence and interchange reconnection. 
The function $F^{\rm FE}$ includes two tunable parameters ($r^{\rm FE}$ and $\sigma^{\rm FE}$), making it crucial to assess the robustness of our findings. 
This appendix explores the impact of these parameters on $\dot{M}_w$.
}

\rev{
In the original $F^{\rm FE}$ formulation given by Eq.~\eqref{eq:energy_flux_emerging_flux}, we assume $\exp[(r^{\rm FE}-r_\ast)/\sigma^{\rm FE}] \gg 1$. 
This holds in our reference case where $r^{\rm FE}/r_\ast = 1.2$ and $\sigma^{\rm FE}/r_\ast = 0.02$. 
However, for smaller $r^{\rm FE}$ or larger $\sigma^{\rm FE}$ values, this may not hold, leading to a reduced $F^{\rm FE}$ at the photosphere. 
To address this, during a parameter survey over $r^{\rm FE}$ and $\sigma^{\rm FE}$, we modify $F^{\rm FE}$ as follows.
\begin{align}
    \frac{F^{\rm FE}}{F^{\rm FE}_{\rm cb}} = \frac{r_{\rm cb}^2 f^{\rm open}_{\rm cb}}{r^2 f^{\rm open}} \frac{\exp \left( \frac{r^{\rm FE}-r}{\sigma^{\rm FE}} \right) \left[ 1 + \exp \left( \frac{r^{\rm FE}-r\ast}{\sigma^{\rm FE}} \right) \right]}{\left[ 1 + \exp \left( \frac{r^{\rm FE}-r}{\sigma^{\rm FE}} \right) \right] \exp \left( \frac{r^{\rm FE}-r_\ast}{\sigma^{\rm FE}} \right)}. \label{eq:energy_flux_emerging_flux_modified}
\end{align}
In the survey, we fix the surface magnetic flux to $\Phi^{\rm surf}/\Phi^{\rm surf}_\odot = 10.86$ and the open-to-surface magnetic flux ratio to $\xi^{\rm open}= 0.1$.
}

\rev{
The top panel of Figure~\ref{fig:Mdot_dependence_on_rloop} illustrates how the mass-loss rate varies with $r^{\rm FE}$. 
The value of $\sigma^{\rm FE}$ is fixed to the reference value ($\sigma^{\rm FE}/r_\ast = 0.02$).
The horizontal dashed line marks the mass-loss rate without flux emergence. 
The mass-loss rate peaks at $r^{\rm FE} = 0.1$, and the reason for this trend is explained as follows.
As $r^{\rm FE}$ increases, more heat is deposited into the supersonic region, which boosts wind velocity rather than mass flux \citep{Leer_1980_JGR, Pneuman_1980_AA}. 
This leads to a decreasing mass-loss rate with larger $r^{\rm FE}$, approaching the no-flux-emergence value. 
The small drop at $r^{\rm FE}<0.1$ likely stems from heating in the chromosphere, which reduces the energy flux transmitted into the corona.
Since the mass-loss rate is nearly proportional to the coronal energy flux \citep{Cranmer_2011_ApJ, Shoda_2020_ApJ}, heat deposition in the chromosphere results in a smaller mass-loss rate.
}

\begin{figure}[t!]
\centering
\includegraphics[width=78mm]{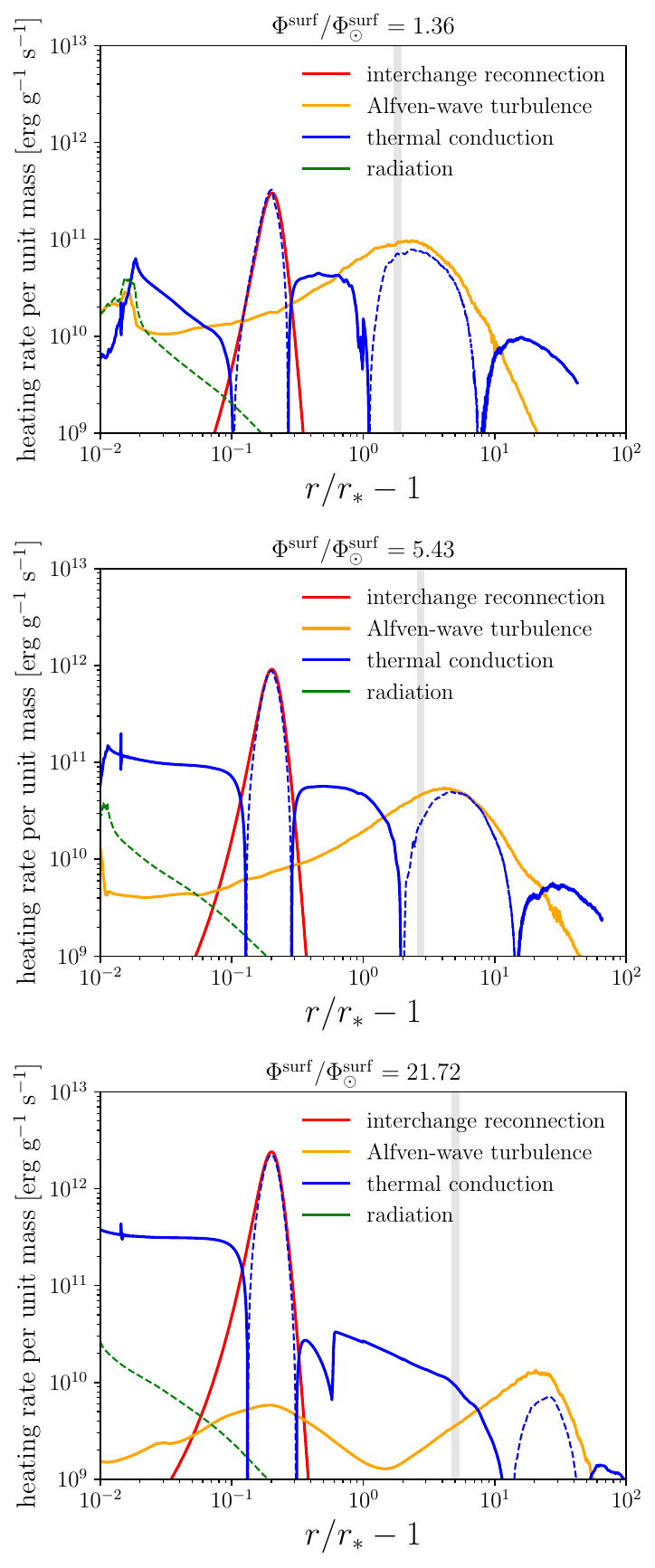}
\caption{
\rev{Radial profiles of heating (thick solid) and cooling (thin dashed) rates per unit mass. 
Mechanisms are color-coded: red for interchange reconnection, orange for Alfv\'en-wave turbulence, green for radiation, and blue for thermal conduction. 
Different panels represent different surface magnetic fluxes: top ($\Phi^{\rm surf}/\Phi^{\rm surf}_\odot = 1.36$), middle ($\Phi^{\rm surf}/\Phi^{\rm surf}_\odot = 5.43$), and bottom ($\Phi^{\rm surf}/\Phi^{\rm surf}_\odot = 21.72$). 
Vertical grey lines indicate sonic points.}
}
\label{fig:heating_rate_comparison}
\end{figure}

\rev{
The bottom panel of Figure~\ref{fig:Mdot_dependence_on_rloop} shows the effect of varying $\sigma^{\rm FE}$, keeping $r^{\rm FE}/r_\ast = 1.2$ constant. 
The mass-loss rate remains constant for $\sigma^{\rm FE}/r_\ast \le 0.1$ and declines for $\sigma^{\rm FE}/r_\ast \ge 0.1$. 
As discussed above, this decrease occurs because heat deposition in the supersonic region becomes larger with increasing $\sigma^{\rm FE}$.
These results confirm the robustness of our results as long as flux emergence heating is limited to the subsonic region.
}

\section{Radial profiles of heating rates \label{app:radial_profiles_of_heating_cooling}}

\rev{
In this appendix, we detail the radial profiles of heating and cooling rates in stellar winds.
Figure~\ref{fig:heating_rate_comparison} displays the radial heating and cooling rates in stellar winds. 
Panels differ by surface magnetic flux: top ($\Phi^{\rm surf}/\Phi^{\rm surf}_\odot = 1.36$), middle ($\Phi^{\rm surf}/\Phi^{\rm surf}_\odot = 5.43$), and bottom ($\Phi^{\rm surf}/\Phi^{\rm surf}_\odot = 21.72$). 
Thick and thin lines represent positive (heating) and negative (cooling) rates, respectively. Colors signify mechanisms: red for interchange reconnection ($Q^{\rm FE}/\rho$), orange for Alfv\'en-wave turbulence ($Q^{\rm turb}/\rho$), blue for thermal conduction ($Q^{\rm cnd}/\rho$), and green for radiation ($Q^{\rm rad}/\rho$). 
We note that Alfv\'en-wave turbulence is the likely dominant dissipation mechanism for Alfv\'en waves in solar wind \citep{Shoda_2018_ApJ_a_self-consistent_model}. 
A grey vertical line in each panel marks the sonic point.
}

\rev{
Regardless of the magnitude of surface magnetic flux, heating by interchange reconnection is localized to the subsonic region, as per Eq.~\eqref{eq:energy_flux_emerging_flux}. 
In contrast, heating by Alfv\'en-wave turbulence is widely distributed and peaks in the supersonic region. 
It is important to clarify that the heating discussed here is in terms of rate per unit mass. 
For larger surface magnetic flux, heating by Alfv\'en-wave turbulence diminishes, whereas heating by interchange reconnection intensifies. 
These trends align with the results in the bottom panel of Figure~\ref{fig:energy_flux_heating_comparison}.
}

\rev{
It is also noteworthy that heating by interchange reconnection is balanced by cooling via thermal conduction. 
Consequently, the lower atmosphere (comprising the coronal base, transition region, and chromosphere) experiences increased conductive heating, particularly when the surface magnetic flux is large, as illustrated in Figure~\ref{fig:heating_rate_comparison}. 
This conductive heating in the chromosphere contributes to coronal mass enhancement through chromospheric evaporation \citep{Yokoyama_2001_ApJ}, making flux emergence and interchange reconnection key factors in elevating solar wind density and mass flux.
}
\end{appendix}

\bibliographystyle{aasjournal}

\begin{thebibliography}{}
    \expandafter\ifx\csname natexlab\endcsname\relax\def\natexlab#1{#1}\fi
    \providecommand{\url}[1]{\href{#1}{#1}}
    \providecommand{\dodoi}[1]{doi:~\href{http://doi.org/#1}{\nolinkurl{#1}}}
    \providecommand{\doeprint}[1]{\href{http://ascl.net/#1}{\nolinkurl{http://ascl.net/#1}}}
    \providecommand{\doarXiv}[1]{\href{https://arxiv.org/abs/#1}{\nolinkurl{https://arxiv.org/abs/#1}}}
    
    \bibitem[{{Abramenko} {et~al.}(2013){Abramenko}, {Zank}, {Dosch}, {Yurchyshyn},
      {Goode}, {Ahn}, \& {Cao}}]{Abramenko_2013_ApJ}
    {Abramenko}, V.~I., {Zank}, G.~P., {Dosch}, A., {et~al.} 2013, \apj, 773, 167,
      \dodoi{10.1088/0004-637X/773/2/167}
    
    \bibitem[{{Airapetian} {et~al.}(2021){Airapetian}, {Jin}, {L{\"u}ftinger},
      {Boro Saikia}, {Kochukhov}, {G{\"u}del}, {Van Der Holst}, \&
      {Manchester}}]{Airapetian_2021_ApJ}
    {Airapetian}, V.~S., {Jin}, M., {L{\"u}ftinger}, T., {et~al.} 2021, \apj, 916,
      96, \dodoi{10.3847/1538-4357/ac081e}
    
    \bibitem[{{Alazraki} \& {Couturier}(1971)}]{Alazraki_1971_AA}
    {Alazraki}, G., \& {Couturier}, P. 1971, \aap, 13, 380
    
    \bibitem[{{Alfv{\'e}n}(1947)}]{Alfven_1947_MNRAS}
    {Alfv{\'e}n}, H. 1947, \mnras, 107, 211, \dodoi{10.1093/mnras/107.2.211}
    
    \bibitem[{{Antiochos} {et~al.}(1999){Antiochos}, {DeVore}, \&
      {Klimchuk}}]{Antiochos_1999_ApJ_A_model_for_solar_coronal_mass_ejection}
    {Antiochos}, S.~K., {DeVore}, C.~R., \& {Klimchuk}, J.~A. 1999, \apj, 510, 485,
      \dodoi{10.1086/306563}
    
    \bibitem[{{Arge} {et~al.}(2023){Arge}, {Leisner}, {Wallace}, \&
      {Henney}}]{Arge_2023_arXiv}
    {Arge}, C.~N., {Leisner}, A., {Wallace}, S., \& {Henney}, C.~J. 2023, arXiv
      e-prints, arXiv:2304.07649, \dodoi{10.48550/arXiv.2304.07649}
    
    \bibitem[{{Arge} \& {Pizzo}(2000)}]{Arge_2000_JGR}
    {Arge}, C.~N., \& {Pizzo}, V.~J. 2000, \jgr, 105, 10465,
      \dodoi{10.1029/1999JA000262}
    
    \bibitem[{{Bale} {et~al.}(2013){Bale}, {Pulupa}, {Salem}, {Chen}, \&
      {Quataert}}]{Bale_2013_ApJ}
    {Bale}, S.~D., {Pulupa}, M., {Salem}, C., {Chen}, C.~H.~K., \& {Quataert}, E.
      2013, \apjl, 769, L22, \dodoi{10.1088/2041-8205/769/2/L22}
    
    \bibitem[{{Bale} {et~al.}(2019){Bale}, {Badman}, {Bonnell}, {Bowen}, {Burgess},
      {Case}, {Cattell}, {Chandran}, {Chaston}, {Chen}, {Drake}, {de Wit},
      {Eastwood}, {Ergun}, {Farrell}, {Fong}, {Goetz}, {Goldstein}, {Goodrich},
      {Harvey}, {Horbury}, {Howes}, {Kasper}, {Kellogg}, {Klimchuk}, {Korreck},
      {Krasnoselskikh}, {Krucker}, {Laker}, {Larson}, {MacDowall}, {Maksimovic},
      {Malaspina}, {Martinez-Oliveros}, {McComas}, {Meyer-Vernet}, {Moncuquet},
      {Mozer}, {Phan}, {Pulupa}, {Raouafi}, {Salem}, {Stansby}, {Stevens}, {Szabo},
      {Velli}, {Woolley}, \& {Wygant}}]{Bale_2019_Nature}
    {Bale}, S.~D., {Badman}, S.~T., {Bonnell}, J.~W., {et~al.} 2019, \nat, 576,
      237, \dodoi{10.1038/s41586-019-1818-7}
    
    \bibitem[{{Bale} {et~al.}(2021){Bale}, {Horbury}, {Velli}, {Desai}, {Halekas},
      {McManus}, {Panasenco}, {Badman}, {Bowen}, {Chandran}, {Drake}, {Kasper},
      {Laker}, {Mallet}, {Matteini}, {Phan}, {Raouafi}, {Squire}, {Woodham}, \&
      {Woolley}}]{Bale_2021_ApJ}
    {Bale}, S.~D., {Horbury}, T.~S., {Velli}, M., {et~al.} 2021, \apj, 923, 174,
      \dodoi{10.3847/1538-4357/ac2d8c}
    
    \bibitem[{{Bale} {et~al.}(2023){Bale}, {Drake}, {McManus}, {Desai}, {Badman},
      {Larson}, {Swisdak}, {Horbury}, {Raouafi}, {Phan}, {Velli}, {McComas},
      {Cohen}, {Mitchell}, {Panasenco}, \& {Kasper}}]{Bale_2023_Nature}
    {Bale}, S.~D., {Drake}, J.~F., {McManus}, M.~D., {et~al.} 2023, \nat, 618, 252,
      \dodoi{10.1038/s41586-023-05955-3}
    
    \bibitem[{{Barnes}(2003)}]{Barnes_2003_ApJ}
    {Barnes}, S.~A. 2003, \apj, 586, 464, \dodoi{10.1086/367639}
    
    \bibitem[{{Belcher}(1971)}]{Belcher_1971_ApJ}
    {Belcher}, J.~W. 1971, \apj, 168, 509, \dodoi{10.1086/151105}
    
    \bibitem[{{Belcher} \& {MacGregor}(1976)}]{Belcher_1976_ApJ}
    {Belcher}, J.~W., \& {MacGregor}, K.~B. 1976, \apj, 210, 498,
      \dodoi{10.1086/154853}
    
    \bibitem[{{Berger} \& {Title}(2001)}]{Berger_2001_ApJ}
    {Berger}, T.~E., \& {Title}, A.~M. 2001, \apj, 553, 449, \dodoi{10.1086/320663}
    
    \bibitem[{{Bourrier} \& {Lecavelier des Etangs}(2013)}]{Bourrier_2013_AA}
    {Bourrier}, V., \& {Lecavelier des Etangs}, A. 2013, \aap, 557, A124,
      \dodoi{10.1051/0004-6361/201321551}
    
    \bibitem[{{Bourrier} {et~al.}(2016){Bourrier}, {Lecavelier des Etangs},
      {Ehrenreich}, {Tanaka}, \& {Vidotto}}]{Bourrier_2016_AA}
    {Bourrier}, V., {Lecavelier des Etangs}, A., {Ehrenreich}, D., {Tanaka}, Y.~A.,
      \& {Vidotto}, A.~A. 2016, \aap, 591, A121,
      \dodoi{10.1051/0004-6361/201628362}
    
    \bibitem[{{Breech} {et~al.}(2008){Breech}, {Matthaeus}, {Minnie}, {Bieber},
      {Oughton}, {Smith}, \& {Isenberg}}]{Breech_2008_JGR}
    {Breech}, B., {Matthaeus}, W.~H., {Minnie}, J., {et~al.} 2008, Journal of
      Geophysical Research (Space Physics), 113, A08105,
      \dodoi{10.1029/2007JA012711}
    
    \bibitem[{{Breu} {et~al.}(2023){Breu}, {Peter}, {Cameron}, \&
      {Solanki}}]{Breu_2023_arXiv}
    {Breu}, C., {Peter}, H., {Cameron}, R., \& {Solanki}, S.~K. 2023, arXiv
      e-prints, arXiv:2305.03769, \dodoi{10.48550/arXiv.2305.03769}
    
    \bibitem[{{Brooks} {et~al.}(2015){Brooks}, {Ugarte-Urra}, \&
      {Warren}}]{Brooks_2015_NatreCommunications}
    {Brooks}, D.~H., {Ugarte-Urra}, I., \& {Warren}, H.~P. 2015, Nature
      Communications, 6, 5947, \dodoi{10.1038/ncomms6947}
    
    \bibitem[{{Brun} {et~al.}(2004){Brun}, {Miesch}, \& {Toomre}}]{Brun_2004_ApJ}
    {Brun}, A.~S., {Miesch}, M.~S., \& {Toomre}, J. 2004, \apj, 614, 1073,
      \dodoi{10.1086/423835}
    
    \bibitem[{{Centeno} {et~al.}(2007){Centeno}, {Socas-Navarro}, {Lites}, {Kubo},
      {Frank}, {Shine}, {Tarbell}, {Title}, {Ichimoto}, {Tsuneta}, {Katsukawa},
      {Suematsu}, {Shimizu}, \& {Nagata}}]{Centeno_2007_ApJ}
    {Centeno}, R., {Socas-Navarro}, H., {Lites}, B., {et~al.} 2007, \apjl, 666,
      L137, \dodoi{10.1086/521726}
    
    \bibitem[{{Chandran} \& {Perez}(2019)}]{Chandran_2019_JPP}
    {Chandran}, B. D.~G., \& {Perez}, J.~C. 2019, Journal of Plasma Physics, 85,
      905850409, \dodoi{10.1017/S0022377819000540}
    
    \bibitem[{{Chebly} {et~al.}(2023){Chebly}, {Alvarado-G{\'o}mez},
      {Poppenh{\"a}ger}, \& {Garraffo}}]{Chebly_2023_MNRAS}
    {Chebly}, J.~J., {Alvarado-G{\'o}mez}, J.~D., {Poppenh{\"a}ger}, K., \&
      {Garraffo}, C. 2023, \mnras, 524, 5060, \dodoi{10.1093/mnras/stad2100}
    
    \bibitem[{{Chitta} {et~al.}(2023{\natexlab{a}}){Chitta}, {Seaton}, {Downs},
      {DeForest}, \& {Higginson}}]{Chitta_2023_NatureAstronomy}
    {Chitta}, L.~P., {Seaton}, D.~B., {Downs}, C., {DeForest}, C.~E., \&
      {Higginson}, A.~K. 2023{\natexlab{a}}, Nature Astronomy, 7, 133,
      \dodoi{10.1038/s41550-022-01834-5}
    
    \bibitem[{{Chitta} {et~al.}(2023{\natexlab{b}}){Chitta}, {Zhukov}, {Berghmans},
      {Peter}, {Parenti}, {Mandal}, {Aznar Cuadrado}, {Sch{\"u}hle}, {Teriaca},
      {Auch{\`e}re}, {Barczynski}, {Buchlin}, {Harra}, {Kraaikamp}, {Long},
      {Rodriguez}, {Schwanitz}, {Smith}, {Verbeeck}, \&
      {Seaton}}]{Chitta_2023_arXiv}
    {Chitta}, L.~P., {Zhukov}, A.~N., {Berghmans}, D., {et~al.} 2023{\natexlab{b}},
      arXiv e-prints, arXiv:2308.13044, \dodoi{10.48550/arXiv.2308.13044}
    
    \bibitem[{{Cohen} {et~al.}(2015){Cohen}, {Ma}, {Drake}, {Glocer}, {Garraffo},
      {Bell}, \& {Gombosi}}]{Cohen_2015_ApJ}
    {Cohen}, O., {Ma}, Y., {Drake}, J.~J., {et~al.} 2015, \apj, 806, 41,
      \dodoi{10.1088/0004-637X/806/1/41}
    
    \bibitem[{{Cranmer}(2010)}]{Cranmer_2010_ApJ_an_efficient_approximation}
    {Cranmer}, S.~R. 2010, \apj, 710, 676, \dodoi{10.1088/0004-637X/710/1/676}
    
    \bibitem[{{Cranmer}(2017)}]{Cranmer_2017_ApJ}
    ---. 2017, \apj, 840, 114, \dodoi{10.3847/1538-4357/aa6f0e}
    
    \bibitem[{{Cranmer} {et~al.}(2021){Cranmer}, {Aarnio}, \&
      {Molnar}}]{Cranmer_2021_AAS}
    {Cranmer}, S.~R., {Aarnio}, A., \& {Molnar}, M.~E. 2021, in American
      Astronomical Society Meeting Abstracts, Vol.~53, American Astronomical
      Society Meeting Abstracts, 131.02
    
    \bibitem[{{Cranmer} \& {Saar}(2011)}]{Cranmer_2011_ApJ}
    {Cranmer}, S.~R., \& {Saar}, S.~H. 2011, \apj, 741, 54,
      \dodoi{10.1088/0004-637X/741/1/54}
    
    \bibitem[{{Cranmer} \& {van Ballegooijen}(2005)}]{Cranmer_2005_ApJ}
    {Cranmer}, S.~R., \& {van Ballegooijen}, A.~A. 2005, \apjs, 156, 265,
      \dodoi{10.1086/426507}
    
    \bibitem[{{Cranmer} \& {van
      Ballegooijen}(2010)}]{Cranmer_2010_ApJ_can_the_solar_wind}
    ---. 2010, \apj, 720, 824, \dodoi{10.1088/0004-637X/720/1/824}
    
    \bibitem[{{Cranmer} \& {van Ballegooijen}(2012)}]{Cranmer_2012_ApJ}
    ---. 2012, \apj, 754, 92, \dodoi{10.1088/0004-637X/754/2/92}
    
    \bibitem[{{Cranmer} {et~al.}(2007){Cranmer}, {van Ballegooijen}, \&
      {Edgar}}]{Cranmer_2007_ApJ}
    {Cranmer}, S.~R., {van Ballegooijen}, A.~A., \& {Edgar}, R.~J. 2007, \apjs,
      171, 520, \dodoi{10.1086/518001}
    
    \bibitem[{{Daly} {et~al.}(2021){Daly}, {Lee}, {Hallis}, {Ishii}, {Bradley},
      {Bland}, {Saxey}, {Fougerouse}, {Rickard}, {Forman}, {Timms}, {Jourdan},
      {Reddy}, {Salge}, {Quadir}, {Christou}, {Cox}, {Aguiar}, {Hattar},
      {Monterrosa}, {Keller}, {Christoffersen}, {Dukes}, {Loeffler}, \&
      {Thompson}}]{Daly_2021_NatAs}
    {Daly}, L., {Lee}, M.~R., {Hallis}, L.~J., {et~al.} 2021, Nature Astronomy, 5,
      1275, \dodoi{10.1038/s41550-021-01487-w}
    
    \bibitem[{{Dere} {et~al.}(1997){Dere}, {Landi}, {Mason}, {Monsignori Fossi}, \&
      {Young}}]{Dere_1997_AA}
    {Dere}, K.~P., {Landi}, E., {Mason}, H.~E., {Monsignori Fossi}, B.~C., \&
      {Young}, P.~R. 1997, \aaps, 125, 149, \dodoi{10.1051/aas:1997368}
    
    \bibitem[{{Dmitruk} {et~al.}(2002){Dmitruk}, {Matthaeus}, {Milano}, {Oughton},
      {Zank}, \& {Mullan}}]{Dmitruk_2002_ApJ}
    {Dmitruk}, P., {Matthaeus}, W.~H., {Milano}, L.~J., {et~al.} 2002, \apj, 575,
      571, \dodoi{10.1086/341188}
    
    \bibitem[{{Drake} {et~al.}(2023){Drake}, {Bale}, {Swisdak}, {Raouafi}, \&
      {Velli}}]{Drake_2023_arXiv}
    {Drake}, J.~F., {Bale}, S.~D., {Swisdak}, M., {Raouafi}, N.~E., \& {Velli}, M.
      2023, arXiv e-prints, arXiv:2306.03425, \dodoi{10.48550/arXiv.2306.03425}
    
    \bibitem[{{Drake} {et~al.}(2021){Drake}, {Agapitov}, {Swisdak}, {Badman},
      {Bale}, {Horbury}, {Kasper}, {MacDowall}, {Mozer}, {Phan}, {Pulupa}, {Szabo},
      \& {Velli}}]{Drake_2021_AA}
    {Drake}, J.~F., {Agapitov}, O., {Swisdak}, M., {et~al.} 2021, \aap, 650, A2,
      \dodoi{10.1051/0004-6361/202039432}
    
    \bibitem[{{Elsasser}(1950)}]{Elsasser_1950_PR}
    {Elsasser}, W.~M. 1950, Physical Review, 79, 183,
      \dodoi{10.1103/PhysRev.79.183}
    
    \bibitem[{{Espagnet} {et~al.}(1995){Espagnet}, {Muller}, {Roudier}, {Mein}, \&
      {Mein}}]{Espagnet_1995_AA}
    {Espagnet}, O., {Muller}, R., {Roudier}, T., {Mein}, N., \& {Mein}, P. 1995,
      \aaps, 109, 79
    
    \bibitem[{{Evensberget} {et~al.}(2021){Evensberget}, {Carter}, {Marsden},
      {Brookshaw}, \& {Folsom}}]{Evensberget_2021_MNRAS}
    {Evensberget}, D., {Carter}, B.~D., {Marsden}, S.~C., {Brookshaw}, L., \&
      {Folsom}, C.~P. 2021, \mnras, 506, 2309, \dodoi{10.1093/mnras/stab1696}
    
    \bibitem[{{Evensberget} {et~al.}(2022){Evensberget}, {Carter}, {Marsden},
      {Brookshaw}, {Folsom}, \& {Salmeron}}]{Evensberget_2022_MNRAS}
    {Evensberget}, D., {Carter}, B.~D., {Marsden}, S.~C., {et~al.} 2022, \mnras,
      510, 5226, \dodoi{10.1093/mnras/stab3557}
    
    \bibitem[{{Evensberget} {et~al.}(2023){Evensberget}, {Marsden}, {Carter},
      {Salmeron}, {Vidotto}, {Folsom}, {Kavanagh}, {Pineda}, {Driessen}, \&
      {Strickert}}]{Evensberget_2023_MNRAS}
    {Evensberget}, D., {Marsden}, S.~C., {Carter}, B.~D., {et~al.} 2023, \mnras,
      524, 2042, \dodoi{10.1093/mnras/stad1650}
    
    \bibitem[{{Fichtinger} {et~al.}(2017){Fichtinger}, {G{\"u}del}, {Mutel},
      {Hallinan}, {Gaidos}, {Skinner}, {Lynch}, \& {Gayley}}]{Fichtinger_2017_AA}
    {Fichtinger}, B., {G{\"u}del}, M., {Mutel}, R.~L., {et~al.} 2017, \aap, 599,
      A127, \dodoi{10.1051/0004-6361/201629886}
    
    \bibitem[{{Finley} {et~al.}(2022){Finley}, {Brun}, {Carlsson}, {Szydlarski},
      {Hansteen}, \& {Shoda}}]{Finley_2022_AA}
    {Finley}, A.~J., {Brun}, A.~S., {Carlsson}, M., {et~al.} 2022, \aap, 665, A118,
      \dodoi{10.1051/0004-6361/202243947}
    
    \bibitem[{{Finley} {et~al.}(2019){Finley}, {Deshmukh}, {Matt}, {Owens}, \&
      {Wu}}]{Finley_2019_ApJ_solar_angular_momentum_loss_past_millenia}
    {Finley}, A.~J., {Deshmukh}, S., {Matt}, S.~P., {Owens}, M., \& {Wu}, C.-J.
      2019, \apj, 883, 67, \dodoi{10.3847/1538-4357/ab3729}
    
    \bibitem[{{Finley} \& {Matt}(2018)}]{Finley_2018_ApJ}
    {Finley}, A.~J., \& {Matt}, S.~P. 2018, \apj, 854, 78,
      \dodoi{10.3847/1538-4357/aaaab5}
    
    \bibitem[{{Fisk}(2003)}]{Fisk_2003_JGR}
    {Fisk}, L.~A. 2003, Journal of Geophysical Research (Space Physics), 108, 1157,
      \dodoi{10.1029/2002JA009284}
    
    \bibitem[{{Fisk} \& {Kasper}(2020)}]{Fisk_2020_ApJ}
    {Fisk}, L.~A., \& {Kasper}, J.~C. 2020, \apjl, 894, L4,
      \dodoi{10.3847/2041-8213/ab8acd}
    
    \bibitem[{{Fisk} {et~al.}(1999){Fisk}, {Schwadron}, \&
      {Zurbuchen}}]{Fisk_1999_JGR}
    {Fisk}, L.~A., {Schwadron}, N.~A., \& {Zurbuchen}, T.~H. 1999, \jgr, 104,
      19765, \dodoi{10.1029/1999JA900256}
    
    \bibitem[{{Folsom} {et~al.}(2016){Folsom}, {Petit}, {Bouvier}, {L{\`e}bre},
      {Amard}, {Palacios}, {Morin}, {Donati}, {Jeffers}, {Marsden}, \&
      {Vidotto}}]{Folsom_2016_MNRAS}
    {Folsom}, C.~P., {Petit}, P., {Bouvier}, J., {et~al.} 2016, \mnras, 457, 580,
      \dodoi{10.1093/mnras/stv2924}
    
    \bibitem[{{Folsom} {et~al.}(2018){Folsom}, {Bouvier}, {Petit}, {L{\`e}bre},
      {Amard}, {Palacios}, {Morin}, {Donati}, \& {Vidotto}}]{Folsom_2018_MNRAS}
    {Folsom}, C.~P., {Bouvier}, J., {Petit}, P., {et~al.} 2018, \mnras, 474, 4956,
      \dodoi{10.1093/mnras/stx3021}
    
    \bibitem[{{Gaidos} {et~al.}(2000){Gaidos}, {G{\"u}del}, \&
      {Blake}}]{Gaidos_2000_GRL}
    {Gaidos}, E.~J., {G{\"u}del}, M., \& {Blake}, G.~A. 2000, \grl, 27, 501,
      \dodoi{10.1029/1999GL010740}
    
    \bibitem[{{Garraffo} {et~al.}(2016){Garraffo}, {Drake}, \&
      {Cohen}}]{Garraffo_2016_AA}
    {Garraffo}, C., {Drake}, J.~J., \& {Cohen}, O. 2016, \aap, 595, A110,
      \dodoi{10.1051/0004-6361/201628367}
    
    \bibitem[{{Garraffo} {et~al.}(2018){Garraffo}, {Drake}, {Dotter}, {Choi},
      {Burke}, {Moschou}, {Alvarado-G{\'o}mez}, {Kashyap}, \&
      {Cohen}}]{Garraffo_2018_ApJ}
    {Garraffo}, C., {Drake}, J.~J., {Dotter}, A., {et~al.} 2018, \apj, 862, 90,
      \dodoi{10.3847/1538-4357/aace5d}
    
    \bibitem[{{Goldstein} {et~al.}(1995){Goldstein}, {Neugebauer}, {Gosling},
      {Bame}, {Phillips}, {McComas}, \& {Balogh}}]{Goldstein_1995_SSRev}
    {Goldstein}, B.~E., {Neugebauer}, M., {Gosling}, J.~T., {et~al.} 1995, \ssr,
      72, 113, \dodoi{10.1007/BF00768764}
    
    \bibitem[{{Goldstein} {et~al.}(1996){Goldstein}, {Neugebauer}, {Phillips},
      {Bame}, {Gosling}, {McComas}, {Wang}, {Sheeley}, \&
      {Suess}}]{Goldstein_1996_AA}
    {Goldstein}, B.~E., {Neugebauer}, M., {Phillips}, J.~L., {et~al.} 1996, \aap,
      316, 296
    
    \bibitem[{{Goodman} \& {Judge}(2012)}]{Goodman_2012_ApJ}
    {Goodman}, M.~L., \& {Judge}, P.~G. 2012, \apj, 751, 75,
      \dodoi{10.1088/0004-637X/751/1/75}
    
    \bibitem[{{Gottlieb} {et~al.}(2001){Gottlieb}, {Shu}, \&
      {Tadmor}}]{Gottlieb_2001_SIAMR}
    {Gottlieb}, S., {Shu}, C.-W., \& {Tadmor}, E. 2001, SIAM Review, 43, 89,
      \dodoi{10.1137/S003614450036757X}
    
    \bibitem[{{G{\"u}del}(2007)}]{Gudel_2007_LRSP}
    {G{\"u}del}, M. 2007, Living Reviews in Solar Physics, 4, 3,
      \dodoi{10.12942/lrsp-2007-3}
    
    \bibitem[{{Gudiksen} \& {Nordlund}(2005)}]{Gudiksen_2005_ApJ}
    {Gudiksen}, B.~V., \& {Nordlund}, {\r{A}}. 2005, \apj, 618, 1020,
      \dodoi{10.1086/426063}
    
    \bibitem[{{Hansteen} \& {Velli}(2012)}]{Hansteen_2012_SSRev}
    {Hansteen}, V.~H., \& {Velli}, M. 2012, \ssr, 172, 89,
      \dodoi{10.1007/s11214-012-9887-z}
    
    \bibitem[{{Harra} {et~al.}(2008){Harra}, {Sakao}, {Mandrini}, {Hara}, {Imada},
      {Young}, {van Driel-Gesztelyi}, \& {Baker}}]{Harra_2008_ApJ}
    {Harra}, L.~K., {Sakao}, T., {Mandrini}, C.~H., {et~al.} 2008, \apjl, 676,
      L147, \dodoi{10.1086/587485}
    
    \bibitem[{{Hollweg}(1986)}]{Hollweg_1986_JGR}
    {Hollweg}, J.~V. 1986, \jgr, 91, 4111, \dodoi{10.1029/JA091iA04p04111}
    
    \bibitem[{{Holzwarth} \& {Jardine}(2007)}]{Holzwarth_2007_AA}
    {Holzwarth}, V., \& {Jardine}, M. 2007, \aap, 463, 11,
      \dodoi{10.1051/0004-6361:20066486}
    
    \bibitem[{{Hossain} {et~al.}(1995){Hossain}, {Gray}, {Pontius}, {Matthaeus}, \&
      {Oughton}}]{Hossain_1995_PhysFluids}
    {Hossain}, M., {Gray}, P.~C., {Pontius}, Duane~H., J., {Matthaeus}, W.~H., \&
      {Oughton}, S. 1995, Physics of Fluids, 7, 2886, \dodoi{10.1063/1.868665}
    
    \bibitem[{{Hotta} {et~al.}(2016){Hotta}, {Rempel}, \&
      {Yokoyama}}]{Hotta_2016_Science}
    {Hotta}, H., {Rempel}, M., \& {Yokoyama}, T. 2016, Science, 351, 1427,
      \dodoi{10.1126/science.aad1893}
    
    \bibitem[{{Hunter}(2007)}]{Hunter_2007_CSE}
    {Hunter}, J.~D. 2007, Computing in Science and Engineering, 9, 90,
      \dodoi{10.1109/MCSE.2007.55}
    
    \bibitem[{{Iijima}(2016)}]{Iijima_2016_PhD}
    {Iijima}, H. 2016, PhD thesis, Department of Earth and Planetary Science,
      School of Science, The University of Tokyo, Japan
    
    \bibitem[{{Iijima} {et~al.}(2023){Iijima}, {Matsumoto}, {Hotta}, \&
      {Imada}}]{Iijima_2023_ApJ}
    {Iijima}, H., {Matsumoto}, T., {Hotta}, H., \& {Imada}, S. 2023, \apjl, 951,
      L47, \dodoi{10.3847/2041-8213/acdde0}
    
    \bibitem[{{Irwin} \& {Bouvier}(2009)}]{Irwin_2009_proceedings}
    {Irwin}, J., \& {Bouvier}, J. 2009, in The Ages of Stars, ed. E.~E. {Mamajek},
      D.~R. {Soderblom}, \& R.~F.~G. {Wyse}, Vol. 258, 363--374,
      \dodoi{10.1017/S1743921309032025}
    
    \bibitem[{{Ishikawa} {et~al.}(2021){Ishikawa}, {Bueno}, {del Pino Alem{\'a}n},
      {Okamoto}, {McKenzie}, {Auch{\`e}re}, {Kano}, {Song}, {Yoshida}, {Rachmeler},
      {Kobayashi}, {Hara}, {Kubo}, {Narukage}, {Sakao}, {Shimizu}, {Suematsu},
      {Bethge}, {De Pontieu}, {Dalda}, {Vigil}, {Winebarger}, {Ballester},
      {Belluzzi}, {{\v{S}}t{\v{e}}p{\'a}n}, {Ramos}, {Carlsson}, \&
      {Leenaarts}}]{Ishikawa_2021_ScienceAdvances}
    {Ishikawa}, R., {Bueno}, J.~T., {del Pino Alem{\'a}n}, T., {et~al.} 2021,
      Science Advances, 7, eabe8406, \dodoi{10.1126/sciadv.abe8406}
    
    \bibitem[{{Ishikawa} {et~al.}(2022){Ishikawa}, {Nakata}, {Katsukawa}, {Masada},
      \& {Riethm{\"u}ller}}]{Ishiwaka_2022_AA}
    {Ishikawa}, R.~T., {Nakata}, M., {Katsukawa}, Y., {Masada}, Y., \&
      {Riethm{\"u}ller}, T.~L. 2022, \aap, 658, A142,
      \dodoi{10.1051/0004-6361/202141743}
    
    \bibitem[{{Jardine} \& {Collier Cameron}(2019)}]{Jardine_2019_MNRAS}
    {Jardine}, M., \& {Collier Cameron}, A. 2019, \mnras, 482, 2853,
      \dodoi{10.1093/mnras/sty2872}
    
    \bibitem[{{Jardine} {et~al.}(2020){Jardine}, {Collier Cameron}, {Donati}, \&
      {Hussain}}]{Jardine_2020_MNRAS}
    {Jardine}, M., {Collier Cameron}, A., {Donati}, J.~F., \& {Hussain}, G.~A.~J.
      2020, \mnras, 491, 4076, \dodoi{10.1093/mnras/stz3173}
    
    \bibitem[{{Johnstone} \& {G{\"u}del}(2015)}]{Johnstone_2015_AA}
    {Johnstone}, C.~P., \& {G{\"u}del}, M. 2015, \aap, 578, A129,
      \dodoi{10.1051/0004-6361/201425283}
    
    \bibitem[{{Judge} {et~al.}(2003){Judge}, {Solomon}, \&
      {Ayres}}]{Judge_2003_ApJ}
    {Judge}, P.~G., {Solomon}, S.~C., \& {Ayres}, T.~R. 2003, \apj, 593, 534,
      \dodoi{10.1086/376405}
    
    \bibitem[{{Kasper} {et~al.}(2019){Kasper}, {Bale}, {Belcher}, {Berthomier},
      {Case}, {Chandran}, {Curtis}, {Gallagher}, {Gary}, {Golub}, {Halekas}, {Ho},
      {Horbury}, {Hu}, {Huang}, {Klein}, {Korreck}, {Larson}, {Livi}, {Maruca},
      {Lavraud}, {Louarn}, {Maksimovic}, {Martinovic}, {McGinnis}, {Pogorelov},
      {Richardson}, {Skoug}, {Steinberg}, {Stevens}, {Szabo}, {Velli},
      {Whittlesey}, {Wright}, {Zank}, {MacDowall}, {McComas}, {McNutt}, {Pulupa},
      {Raouafi}, \& {Schwadron}}]{Kasper_2019_Nature}
    {Kasper}, J.~C., {Bale}, S.~D., {Belcher}, J.~W., {et~al.} 2019, \nat, 576,
      228, \dodoi{10.1038/s41586-019-1813-z}
    
    \bibitem[{{Kavanagh} {et~al.}(2019){Kavanagh}, {Vidotto}, {{\'O}.
      Fionnag{\'a}in}, {Bourrier}, {Fares}, {Jardine}, {Helling}, {Moutou},
      {Llama}, \& {Wheatley}}]{Kavanagh_2019_MNRAS}
    {Kavanagh}, R.~D., {Vidotto}, A.~A., {{\'O}. Fionnag{\'a}in}, D., {et~al.}
      2019, \mnras, 485, 4529, \dodoi{10.1093/mnras/stz655}
    
    \bibitem[{{Kawaler}(1988)}]{Kawaler_1988_ApJ}
    {Kawaler}, S.~D. 1988, \apj, 333, 236, \dodoi{10.1086/166740}
    
    \bibitem[{{Kislyakova} {et~al.}(2013){Kislyakova}, {Lammer}, {Holmstr{\"o}m},
      {Panchenko}, {Odert}, {Erkaev}, {Leitzinger}, {Khodachenko}, {Kulikov},
      {G{\"u}del}, \& {Hanslmeier}}]{Kislyakova_2013_AsBio}
    {Kislyakova}, K.~G., {Lammer}, H., {Holmstr{\"o}m}, M., {et~al.} 2013,
      Astrobiology, 13, 1030, \dodoi{10.1089/ast.2012.0958}
    
    \bibitem[{{Kislyakova} {et~al.}(2019){Kislyakova}, {Holmstr{\"o}m}, {Odert},
      {Lammer}, {Erkaev}, {Khodachenko}, {Shaikhislamov}, {Dorfi}, \&
      {G{\"u}del}}]{Kislyakova_2019_AA}
    {Kislyakova}, K.~G., {Holmstr{\"o}m}, M., {Odert}, P., {et~al.} 2019, \aap,
      623, A131, \dodoi{10.1051/0004-6361/201833941}
    
    \bibitem[{{Kochukhov} {et~al.}(2020){Kochukhov}, {Hackman}, {Lehtinen}, \&
      {Wehrhahn}}]{Kochukhov_2020_AA}
    {Kochukhov}, O., {Hackman}, T., {Lehtinen}, J.~J., \& {Wehrhahn}, A. 2020,
      \aap, 635, A142, \dodoi{10.1051/0004-6361/201937185}
    
    \bibitem[{{Kopp} \& {Holzer}(1976)}]{Kopp_1976_SolPhys}
    {Kopp}, R.~A., \& {Holzer}, T.~E. 1976, \solphys, 49, 43,
      \dodoi{10.1007/BF00221484}
    
    \bibitem[{{Kostik} {et~al.}(2009){Kostik}, {Khomenko}, \&
      {Shchukina}}]{Kostik_2009_AA}
    {Kostik}, R., {Khomenko}, E., \& {Shchukina}, N. 2009, \aap, 506, 1405,
      \dodoi{10.1051/0004-6361/200912441}
    
    \bibitem[{{Kraft}(1967)}]{Kraft_1967_ApJ}
    {Kraft}, R.~P. 1967, \apj, 150, 551, \dodoi{10.1086/149359}
    
    \bibitem[{{Kumar} {et~al.}(2023){Kumar}, {Karpen}, {Uritsky}, {Deforest},
      {Raouafi}, {DeVore}, \& {Antiochos}}]{Kumar_2023_ApJ}
    {Kumar}, P., {Karpen}, J.~T., {Uritsky}, V.~M., {et~al.} 2023, \apjl, 951, L15,
      \dodoi{10.3847/2041-8213/acd54e}
    
    \bibitem[{{Kuniyoshi} {et~al.}(2023){Kuniyoshi}, {Shoda}, {Iijima}, \&
      {Yokoyama}}]{Kuniyoshi_2023_ApJ}
    {Kuniyoshi}, H., {Shoda}, M., {Iijima}, H., \& {Yokoyama}, T. 2023, \apj, 949,
      8, \dodoi{10.3847/1538-4357/accbb8}
    
    \bibitem[{{Landi} {et~al.}(2012){Landi}, {Del Zanna}, {Young}, {Dere}, \&
      {Mason}}]{Landi_2012_ApJ}
    {Landi}, E., {Del Zanna}, G., {Young}, P.~R., {Dere}, K.~P., \& {Mason}, H.~E.
      2012, \apj, 744, 99, \dodoi{10.1088/0004-637X/744/2/99}
    
    \bibitem[{{Le Chat} {et~al.}(2012){Le Chat}, {Issautier}, \&
      {Meyer-Vernet}}]{Le_Chat_2012_SolPhys}
    {Le Chat}, G., {Issautier}, K., \& {Meyer-Vernet}, N. 2012, \solphys, 279, 197,
      \dodoi{10.1007/s11207-012-9967-y}
    
    \bibitem[{{Leer} \& {Holzer}(1980)}]{Leer_1980_JGR}
    {Leer}, E., \& {Holzer}, T.~E. 1980, \jgr, 85, 4681,
      \dodoi{10.1029/JA085iA09p04681}
    
    \bibitem[{{Lim} \& {White}(1996)}]{Lim_1996_ApJ_outflow_from_late_type_stars}
    {Lim}, J., \& {White}, S.~M. 1996, \apjl, 462, L91, \dodoi{10.1086/310038}
    
    \bibitem[{{Lim} {et~al.}(1996){Lim}, {White}, \&
      {Slee}}]{Lim_1996_ApJ_Mdot_of_ProxCen}
    {Lim}, J., {White}, S.~M., \& {Slee}, O.~B. 1996, \apj, 460, 976,
      \dodoi{10.1086/177025}
    
    \bibitem[{{Linker} {et~al.}(2017){Linker}, {Caplan}, {Downs}, {Riley}, {Mikic},
      {Lionello}, {Henney}, {Arge}, {Liu}, {Derosa}, {Yeates}, \&
      {Owens}}]{Linker_2017_ApJ}
    {Linker}, J.~A., {Caplan}, R.~M., {Downs}, C., {et~al.} 2017, \apj, 848, 70,
      \dodoi{10.3847/1538-4357/aa8a70}
    
    \bibitem[{{Linker} {et~al.}(2021){Linker}, {Heinemann}, {Temmer}, {Owens},
      {Caplan}, {Arge}, {Asvestari}, {Delouille}, {Downs}, {Hofmeister}, {Jebaraj},
      {Madjarska}, {Pinto}, {Pomoell}, {Samara}, {Scolini}, \&
      {Vr{\v{s}}nak}}]{Linker_2021_ApJ}
    {Linker}, J.~A., {Heinemann}, S.~G., {Temmer}, M., {et~al.} 2021, \apj, 918,
      21, \dodoi{10.3847/1538-4357/ac090a}
    
    \bibitem[{{Lionello} {et~al.}(2016){Lionello}, {T{\"o}r{\"o}k}, {Titov},
      {Leake}, {Miki{\'c}}, {Linker}, \& {Linton}}]{Lionello_2016_ApJ}
    {Lionello}, R., {T{\"o}r{\"o}k}, T., {Titov}, V.~S., {et~al.} 2016, \apjl, 831,
      L2, \dodoi{10.3847/2041-8205/831/1/L2}
    
    \bibitem[{{Magaudda} {et~al.}(2020){Magaudda}, {Stelzer}, {Covey}, {Raetz},
      {Matt}, \& {Scholz}}]{Magaudda_2020_AA}
    {Magaudda}, E., {Stelzer}, B., {Covey}, K.~R., {et~al.} 2020, \aap, 638, A20,
      \dodoi{10.1051/0004-6361/201937408}
    
    \bibitem[{{Magyar} \& {Nakariakov}(2021)}]{Magyar_2021_ApJ_GPM_solar_wind}
    {Magyar}, N., \& {Nakariakov}, V.~M. 2021, \apj, 907, 55,
      \dodoi{10.3847/1538-4357/abd02f}
    
    \bibitem[{{Magyar} {et~al.}(2021){Magyar}, {Utz}, {Erd{\'e}lyi}, \&
      {Nakariakov}}]{Magyar_2021_ApJ_switchback_II}
    {Magyar}, N., {Utz}, D., {Erd{\'e}lyi}, R., \& {Nakariakov}, V.~M. 2021, \apj,
      914, 8, \dodoi{10.3847/1538-4357/abfa98}
    
    \bibitem[{{Masuda} {et~al.}(1994){Masuda}, {Kosugi}, {Hara}, {Tsuneta}, \&
      {Ogawara}}]{Masuda_1994_Nature}
    {Masuda}, S., {Kosugi}, T., {Hara}, H., {Tsuneta}, S., \& {Ogawara}, Y. 1994,
      \nat, 371, 495, \dodoi{10.1038/371495a0}
    
    \bibitem[{{Matsumoto}(2021)}]{Matsumoto_2021_MNRAS}
    {Matsumoto}, T. 2021, \mnras, 500, 4779, \dodoi{10.1093/mnras/staa3533}
    
    \bibitem[{{Matsumoto} \& {Shibata}(2010)}]{Matsumoto_2010_ApJ}
    {Matsumoto}, T., \& {Shibata}, K. 2010, \apj, 710, 1857,
      \dodoi{10.1088/0004-637X/710/2/1857}
    
    \bibitem[{{Matt} {et~al.}(2012){Matt}, {MacGregor}, {Pinsonneault}, \&
      {Greene}}]{Matt_2012_ApJ}
    {Matt}, S.~P., {MacGregor}, K.~B., {Pinsonneault}, M.~H., \& {Greene}, T.~P.
      2012, \apjl, 754, L26, \dodoi{10.1088/2041-8205/754/2/L26}
    
    \bibitem[{{Matthaeus} {et~al.}(1994){Matthaeus}, {Oughton}, {Pontius}, \&
      {Zhou}}]{Matthaeus_1994_ApJ}
    {Matthaeus}, W.~H., {Oughton}, S., {Pontius}, Duane~H., J., \& {Zhou}, Y. 1994,
      \jgr, 99, 19267, \dodoi{10.1029/94JA01233}
    
    \bibitem[{{Matthaeus} {et~al.}(1999){Matthaeus}, {Zank}, {Oughton}, {Mullan},
      \& {Dmitruk}}]{Matthaeus_1999_ApJ}
    {Matthaeus}, W.~H., {Zank}, G.~P., {Oughton}, S., {Mullan}, D.~J., \&
      {Dmitruk}, P. 1999, \apjl, 523, L93, \dodoi{10.1086/312259}
    
    \bibitem[{{McCann} {et~al.}(2019){McCann}, {Murray-Clay}, {Kratter}, \&
      {Krumholz}}]{McCann_2019_ApJ}
    {McCann}, J., {Murray-Clay}, R.~A., {Kratter}, K., \& {Krumholz}, M.~R. 2019,
      \apj, 873, 89, \dodoi{10.3847/1538-4357/ab05b8}
    
    \bibitem[{{Mestel}(1968)}]{Mestel_1968_MNRAS}
    {Mestel}, L. 1968, \mnras, 138, 359, \dodoi{10.1093/mnras/138.3.359}
    
    \bibitem[{{Metcalfe} {et~al.}(2023){Metcalfe}, {Strassmeier}, {Ilyin}, {van
      Saders}, {Ayres}, {Finley}, {Kochukhov}, {Petit}, {See}, {Stassun},
      {Jeffers}, {Marsden}, {Morin}, \& {Vidotto}}]{Metcalfe_2023_ApJ}
    {Metcalfe}, T.~S., {Strassmeier}, K.~G., {Ilyin}, I.~V., {et~al.} 2023, \apjl,
      948, L6, \dodoi{10.3847/2041-8213/acce38}
    
    \bibitem[{{Meyer} {et~al.}(2012){Meyer}, {Balsara}, \&
      {Aslam}}]{Meyer_2012_MNRAS}
    {Meyer}, C.~D., {Balsara}, D.~S., \& {Aslam}, T.~D. 2012, \mnras, 422, 2102,
      \dodoi{10.1111/j.1365-2966.2012.20744.x}
    
    \bibitem[{{Meyer} {et~al.}(2014){Meyer}, {Balsara}, \&
      {Aslam}}]{Meyer_2014_JCP}
    ---. 2014, Journal of Computational Physics, 257, 594,
      \dodoi{10.1016/j.jcp.2013.08.021}
    
    \bibitem[{{Mitani} {et~al.}(2022){Mitani}, {Nakatani}, \&
      {Yoshida}}]{Mitani_2022_MNRAS}
    {Mitani}, H., {Nakatani}, R., \& {Yoshida}, N. 2022, \mnras, 512, 855,
      \dodoi{10.1093/mnras/stac556}
    
    \bibitem[{{Miyoshi} \& {Kusano}(2005)}]{Miyoshi_2005_JCP}
    {Miyoshi}, T., \& {Kusano}, K. 2005, Journal of Computational Physics, 208,
      315, \dodoi{10.1016/j.jcp.2005.02.017}
    
    \bibitem[{{Morton} {et~al.}(2019){Morton}, {Weberg}, \&
      {McLaughlin}}]{Morton_2019_Nature_Astronomy}
    {Morton}, R.~J., {Weberg}, M.~J., \& {McLaughlin}, J.~A. 2019, Nature
      Astronomy, 3, 223, \dodoi{10.1038/s41550-018-0668-9}
    
    \bibitem[{{Namekata} {et~al.}(2017){Namekata}, {Sakaue}, {Watanabe}, {Asai},
      {Maehara}, {Notsu}, {Notsu}, {Honda}, {Ishii}, {Ikuta}, {Nogami}, \&
      {Shibata}}]{Namekata_2017_ApJ}
    {Namekata}, K., {Sakaue}, T., {Watanabe}, K., {et~al.} 2017, \apj, 851, 91,
      \dodoi{10.3847/1538-4357/aa9b34}
    
    \bibitem[{{Namekata} {et~al.}(2021){Namekata}, {Maehara}, {Honda}, {Notsu},
      {Okamoto}, {Takahashi}, {Takayama}, {Ohshima}, {Saito}, {Katoh}, {Tozuka},
      {Murata}, {Ogawa}, {Niwano}, {Adachi}, {Oeda}, {Shiraishi}, {Isogai}, {Seki},
      {Ishii}, {Ichimoto}, {Nogami}, \& {Shibata}}]{Namekata_2021_NatAs}
    {Namekata}, K., {Maehara}, H., {Honda}, S., {et~al.} 2021, Nature Astronomy, 6,
      241, \dodoi{10.1038/s41550-021-01532-8}
    
    \bibitem[{{{\'O} Fionnag{\'a}in} {et~al.}(2019){{\'O} Fionnag{\'a}in},
      {Vidotto}, {Petit}, {Folsom}, {Jeffers}, {Marsden}, {Morin}, {do Nascimento},
      \& {BCool Collaboration}}]{OFionnagain_2019_MNRAS}
    {{\'O} Fionnag{\'a}in}, D., {Vidotto}, A.~A., {Petit}, P., {et~al.} 2019,
      \mnras, 483, 873, \dodoi{10.1093/mnras/sty3132}
    
    \bibitem[{{Ofman}(2004)}]{Ofman_2004_JGRA}
    {Ofman}, L. 2004, Journal of Geophysical Research (Space Physics), 109, A07102,
      \dodoi{10.1029/2003JA010221}
    
    \bibitem[{{Ofman} \& {Davila}(1995)}]{Ofman_1995_JGR}
    {Ofman}, L., \& {Davila}, J.~M. 1995, \jgr, 100, 23413,
      \dodoi{10.1029/95JA02222}
    
    \bibitem[{{Ofman} \& {Davila}(1998)}]{Ofman_1998_JGR}
    ---. 1998, \jgr, 103, 23677, \dodoi{10.1029/98JA01996}
    
    \bibitem[{{Okamoto} \& {Sakurai}(2018)}]{Okamoto_2018_ApJ}
    {Okamoto}, T.~J., \& {Sakurai}, T. 2018, \apjl, 852, L16,
      \dodoi{10.3847/2041-8213/aaa3d8}
    
    \bibitem[{{Osterbrock}(1961)}]{Osterbrock_1961_ApJ}
    {Osterbrock}, D.~E. 1961, \apj, 134, 347, \dodoi{10.1086/147165}
    
    \bibitem[{{Pallavicini} {et~al.}(1981){Pallavicini}, {Golub}, {Rosner},
      {Vaiana}, {Ayres}, \& {Linsky}}]{Pallavicini_1981_ApJ}
    {Pallavicini}, R., {Golub}, L., {Rosner}, R., {et~al.} 1981, \apj, 248, 279,
      \dodoi{10.1086/159152}
    
    \bibitem[{{Panasenco} {et~al.}(2019){Panasenco}, {Velli}, \&
      {Panasenco}}]{Panasenco_2019_ApJ}
    {Panasenco}, O., {Velli}, M., \& {Panasenco}, A. 2019, \apj, 873, 25,
      \dodoi{10.3847/1538-4357/ab017c}
    
    \bibitem[{{Parker}(1955)}]{Parker_1955_ApJ}
    {Parker}, E.~N. 1955, \apj, 122, 293, \dodoi{10.1086/146087}
    
    \bibitem[{{Parker}(1958)}]{Parker_1958_ApJ}
    ---. 1958, \apj, 128, 664, \dodoi{10.1086/146579}
    
    \bibitem[{{Parker}(1983)}]{Parker_1983_ApJ}
    ---. 1983, \apj, 264, 642, \dodoi{10.1086/160637}
    
    \bibitem[{{Paxton} {et~al.}(2011){Paxton}, {Bildsten}, {Dotter}, {Herwig},
      {Lesaffre}, \& {Timmes}}]{Paxton_2011_ApJS}
    {Paxton}, B., {Bildsten}, L., {Dotter}, A., {et~al.} 2011, \apjs, 192, 3,
      \dodoi{10.1088/0067-0049/192/1/3}
    
    \bibitem[{{Peres} {et~al.}(2000){Peres}, {Orlando}, {Reale}, {Rosner}, \&
      {Hudson}}]{Peres_2000_ApJ}
    {Peres}, G., {Orlando}, S., {Reale}, F., {Rosner}, R., \& {Hudson}, H. 2000,
      \apj, 528, 537, \dodoi{10.1086/308136}
    
    \bibitem[{{Perez} \& {Chandran}(2013)}]{Perez_2013_ApJ}
    {Perez}, J.~C., \& {Chandran}, B. D.~G. 2013, \apj, 776, 124,
      \dodoi{10.1088/0004-637X/776/2/124}
    
    \bibitem[{{Pevtsov} {et~al.}(2003){Pevtsov}, {Fisher}, {Acton}, {Longcope},
      {Johns-Krull}, {Kankelborg}, \& {Metcalf}}]{Pevtsov_2003_ApJ}
    {Pevtsov}, A.~A., {Fisher}, G.~H., {Acton}, L.~W., {et~al.} 2003, \apj, 598,
      1387, \dodoi{10.1086/378944}
    
    \bibitem[{{Pizzolato} {et~al.}(2003){Pizzolato}, {Maggio}, {Micela},
      {Sciortino}, \& {Ventura}}]{Pizzolato_2003_AA}
    {Pizzolato}, N., {Maggio}, A., {Micela}, G., {Sciortino}, S., \& {Ventura}, P.
      2003, \aap, 397, 147, \dodoi{10.1051/0004-6361:20021560}
    
    \bibitem[{{Pneuman}(1980)}]{Pneuman_1980_AA}
    {Pneuman}, G.~W. 1980, \aap, 81, 161
    
    \bibitem[{{Raouafi} {et~al.}(2023){Raouafi}, {Stenborg}, {Seaton}, {Wang},
      {Wang}, {DeForest}, {Bale}, {Drake}, {Uritsky}, {Karpen}, {DeVore},
      {Sterling}, {Horbury}, {Harra}, {Bourouaine}, {Kasper}, {Kumar}, {Phan}, \&
      {Velli}}]{Raouafi_2023_ApJ}
    {Raouafi}, N.~E., {Stenborg}, G., {Seaton}, D.~B., {et~al.} 2023, \apj, 945,
      28, \dodoi{10.3847/1538-4357/acaf6c}
    
    \bibitem[{{Rappazzo} {et~al.}(2012){Rappazzo}, {Matthaeus}, {Ruffolo},
      {Servidio}, \& {Velli}}]{Rappazzo_2012_ApJ}
    {Rappazzo}, A.~F., {Matthaeus}, W.~H., {Ruffolo}, D., {Servidio}, S., \&
      {Velli}, M. 2012, \apjl, 758, L14, \dodoi{10.1088/2041-8205/758/1/L14}
    
    \bibitem[{{Reames}(1999)}]{Reames_1999_SSRev}
    {Reames}, D.~V. 1999, \ssr, 90, 413, \dodoi{10.1023/A:1005105831781}
    
    \bibitem[{{Reiners} {et~al.}(2009){Reiners}, {Basri}, \&
      {Browning}}]{Reiners_2009_ApJ}
    {Reiners}, A., {Basri}, G., \& {Browning}, M. 2009, \apj, 692, 538,
      \dodoi{10.1088/0004-637X/692/1/538}
    
    \bibitem[{{Reiners} {et~al.}(2022){Reiners}, {Shulyak}, {K{\"a}pyl{\"a}},
      {Ribas}, {Nagel}, {Zechmeister}, {Caballero}, {Shan}, {Fuhrmeister},
      {Quirrenbach}, {Amado}, {Montes}, {Jeffers}, {Azzaro}, {B{\'e}jar},
      {Chaturvedi}, {Henning}, {K{\"u}rster}, \& {Pall{\'e}}}]{Reiners_2022_AA}
    {Reiners}, A., {Shulyak}, D., {K{\"a}pyl{\"a}}, P.~J., {et~al.} 2022, \aap,
      662, A41, \dodoi{10.1051/0004-6361/202243251}
    
    \bibitem[{{R{\'e}ville} {et~al.}(2015){R{\'e}ville}, {Brun}, {Matt},
      {Strugarek}, \& {Pinto}}]{Reville_2015_ApJ}
    {R{\'e}ville}, V., {Brun}, A.~S., {Matt}, S.~P., {Strugarek}, A., \& {Pinto},
      R.~F. 2015, \apj, 798, 116, \dodoi{10.1088/0004-637X/798/2/116}
    
    \bibitem[{{R{\'e}ville} {et~al.}(2020){R{\'e}ville}, {Velli}, {Panasenco},
      {Tenerani}, {Shi}, {Badman}, {Bale}, {Kasper}, {Stevens}, {Korreck},
      {Bonnell}, {Case}, {de Wit}, {Goetz}, {Harvey}, {Larson}, {Livi},
      {Malaspina}, {MacDowall}, {Pulupa}, \& {Whittlesey}}]{Reville_2020_ApJ}
    {R{\'e}ville}, V., {Velli}, M., {Panasenco}, O., {et~al.} 2020, \apjs, 246, 24,
      \dodoi{10.3847/1538-4365/ab4fef}
    
    \bibitem[{{Riley} {et~al.}(2015){Riley}, {Linker}, \&
      {Arge}}]{Riley_2015_SpaceWeather}
    {Riley}, P., {Linker}, J.~A., \& {Arge}, C.~N. 2015, Space Weather, 13, 154,
      \dodoi{10.1002/2014SW001144}
    
    \bibitem[{{Saar}(1996)}]{Saar_1996_proceedings}
    {Saar}, S.~H. 1996, in Stellar Surface Structure, ed. K.~G. {Strassmeier} \&
      J.~L. {Linsky}, Vol. 176, 237
    
    \bibitem[{{Saar}(2001)}]{Saar_2001_proceedings}
    {Saar}, S.~H. 2001, in Astronomical Society of the Pacific Conference Series,
      Vol. 223, 11th Cambridge Workshop on Cool Stars, Stellar Systems and the Sun,
      ed. R.~J. {Garcia Lopez}, R.~{Rebolo}, \& M.~R. {Zapaterio Osorio}, 292
    
    \bibitem[{{Sakao} {et~al.}(2007){Sakao}, {Kano}, {Narukage}, {Kotoku}, {Bando},
      {DeLuca}, {Lundquist}, {Tsuneta}, {Harra}, {Katsukawa}, {Kubo}, {Hara},
      {Matsuzaki}, {Shimojo}, {Bookbinder}, {Golub}, {Korreck}, {Su}, {Shibasaki},
      {Shimizu}, \& {Nakatani}}]{Sakao_2007_Science}
    {Sakao}, T., {Kano}, R., {Narukage}, N., {et~al.} 2007, Science, 318, 1585,
      \dodoi{10.1126/science.1147292}
    
    \bibitem[{{Sakurai}(1985)}]{Sakurai_1985_AA}
    {Sakurai}, T. 1985, \aap, 152, 121
    
    \bibitem[{{Salem} {et~al.}(2003){Salem}, {Hubert}, {Lacombe}, {Bale},
      {Mangeney}, {Larson}, \& {Lin}}]{Salem_2003_ApJ}
    {Salem}, C., {Hubert}, D., {Lacombe}, C., {et~al.} 2003, \apj, 585, 1147,
      \dodoi{10.1086/346185}
    
    \bibitem[{{Salucci} {et~al.}(1994){Salucci}, {Bertello}, {Cavallini},
      {Ceppatelli}, \& {Righini}}]{Salucci_1994_AA}
    {Salucci}, G., {Bertello}, L., {Cavallini}, F., {Ceppatelli}, G., \& {Righini},
      A. 1994, \aap, 285, 322
    
    \bibitem[{{Sanz-Forcada} {et~al.}(2011){Sanz-Forcada}, {Micela}, {Ribas},
      {Pollock}, {Eiroa}, {Velasco}, {Solano}, \&
      {Garc{\'\i}a-{\'A}lvarez}}]{Sanz-Forcada_2011_AA}
    {Sanz-Forcada}, J., {Micela}, G., {Ribas}, I., {et~al.} 2011, \aap, 532, A6,
      \dodoi{10.1051/0004-6361/201116594}
    
    \bibitem[{{Schleich} {et~al.}(2023){Schleich}, {Boro Saikia}, {Ziegler},
      {G{\"u}del}, \& {Bartel}}]{Schleich_2023_AA}
    {Schleich}, S., {Boro Saikia}, S., {Ziegler}, U., {G{\"u}del}, M., \& {Bartel},
      M. 2023, \aap, 672, A64, \dodoi{10.1051/0004-6361/202245009}
    
    \bibitem[{{Schr{\"o}der} \& {Cuntz}(2005)}]{Schroder_2005_ApJ}
    {Schr{\"o}der}, K.~P., \& {Cuntz}, M. 2005, \apjl, 630, L73,
      \dodoi{10.1086/491579}
    
    \bibitem[{{Schwadron} \& {McComas}(2021)}]{Schwadron_2021_ApJ}
    {Schwadron}, N.~A., \& {McComas}, D.~J. 2021, \apj, 909, 95,
      \dodoi{10.3847/1538-4357/abd4e6}
    
    \bibitem[{{See} {et~al.}(2019){See}, {Matt}, {Folsom}, {Boro Saikia}, {Donati},
      {Fares}, {Finley}, {H{\'e}brard}, {Jardine}, {Jeffers}, {Lehmann}, {Marsden},
      {Mengel}, {Morin}, {Petit}, {Vidotto}, {Waite}, \& {BCool
      Collaboration}}]{See_2019_ApJ}
    {See}, V., {Matt}, S.~P., {Folsom}, C.~P., {et~al.} 2019, \apj, 876, 118,
      \dodoi{10.3847/1538-4357/ab1096}
    
    \bibitem[{{Sharma} \& {Morton}(2023)}]{Sharma_2023_NatAs}
    {Sharma}, R., \& {Morton}, R.~J. 2023, Nature Astronomy,
      \dodoi{10.1038/s41550-023-02070-1}
    
    \bibitem[{{Shibata} \& {Magara}(2011)}]{Shibata_2011_LRSP}
    {Shibata}, K., \& {Magara}, T. 2011, Living Reviews in Solar Physics, 8, 6,
      \dodoi{10.12942/lrsp-2011-6}
    
    \bibitem[{{Shoda} {et~al.}(2021){Shoda}, {Chandran}, \&
      {Cranmer}}]{Shoda_2021_ApJ}
    {Shoda}, M., {Chandran}, B. D.~G., \& {Cranmer}, S.~R. 2021, \apj, 915, 52,
      \dodoi{10.3847/1538-4357/abfdbc}
    
    \bibitem[{{Shoda} {et~al.}(2019){Shoda}, {Suzuki}, {Asgari-Targhi}, \&
      {Yokoyama}}]{Shoda_2019_ApJ}
    {Shoda}, M., {Suzuki}, T.~K., {Asgari-Targhi}, M., \& {Yokoyama}, T. 2019,
      \apjl, 880, L2, \dodoi{10.3847/2041-8213/ab2b45}
    
    \bibitem[{{Shoda} \& {Takasao}(2021)}]{Shoda_2021_AA}
    {Shoda}, M., \& {Takasao}, S. 2021, \aap, 656, A111,
      \dodoi{10.1051/0004-6361/202141563}
    
    \bibitem[{{Shoda} {et~al.}(2018){Shoda}, {Yokoyama}, \&
      {Suzuki}}]{Shoda_2018_ApJ_a_self-consistent_model}
    {Shoda}, M., {Yokoyama}, T., \& {Suzuki}, T.~K. 2018, \apj, 853, 190,
      \dodoi{10.3847/1538-4357/aaa3e1}
    
    \bibitem[{{Shoda} {et~al.}(2020){Shoda}, {Suzuki}, {Matt}, {Cranmer},
      {Vidotto}, {Strugarek}, {See}, {R{\'e}ville}, {Finley}, \&
      {Brun}}]{Shoda_2020_ApJ}
    {Shoda}, M., {Suzuki}, T.~K., {Matt}, S.~P., {et~al.} 2020, \apj, 896, 123,
      \dodoi{10.3847/1538-4357/ab94bf}
    
    \bibitem[{{Shu} \& {Osher}(1988)}]{Shu_1988_JCP}
    {Shu}, C.-W., \& {Osher}, S. 1988, Journal of Computational Physics, 77, 439,
      \dodoi{10.1016/0021-9991(88)90177-5}
    
    \bibitem[{{Skumanich}(1972)}]{Skumanich_1972_ApJ}
    {Skumanich}, A. 1972, \apj, 171, 565, \dodoi{10.1086/151310}
    
    \bibitem[{{Smith} \& {Balogh}(1995)}]{Smith_1995_GRL}
    {Smith}, E.~J., \& {Balogh}, A. 1995, \grl, 22, 3317, \dodoi{10.1029/95GL02826}
    
    \bibitem[{{Spitzer} \& {H{\"a}rm}(1953)}]{Spitzer_1953_PhysRev}
    {Spitzer}, L., \& {H{\"a}rm}, R. 1953, Physical Review, 89, 977,
      \dodoi{10.1103/PhysRev.89.977}
    
    \bibitem[{{Squire} {et~al.}(2020){Squire}, {Chandran}, \&
      {Meyrand}}]{Squire_2020_ApJ}
    {Squire}, J., {Chandran}, B.~D.~G., \& {Meyrand}, R. 2020, \apjl, 891, L2,
      \dodoi{10.3847/2041-8213/ab74e1}
    
    \bibitem[{{Suresh} \& {Huynh}(1997)}]{Suresh_1997_JCP}
    {Suresh}, A., \& {Huynh}, H.~T. 1997, Journal of Computational Physics, 136,
      83, \dodoi{10.1006/jcph.1997.5745}
    
    \bibitem[{{Suzuki}(2018)}]{Suzuki_2018_PASJ}
    {Suzuki}, T.~K. 2018, \pasj, 70, 34, \dodoi{10.1093/pasj/psy023}
    
    \bibitem[{{Suzuki} {et~al.}(2013){Suzuki}, {Imada}, {Kataoka}, {Kato},
      {Matsumoto}, {Miyahara}, \& {Tsuneta}}]{Suzuki_2013_PASJ}
    {Suzuki}, T.~K., {Imada}, S., {Kataoka}, R., {et~al.} 2013, \pasj, 65, 98,
      \dodoi{10.1093/pasj/65.5.98}
    
    \bibitem[{{Suzuki} \& {Inutsuka}(2005)}]{Suzuki_2005_ApJ}
    {Suzuki}, T.~K., \& {Inutsuka}, S.-i. 2005, \apjl, 632, L49,
      \dodoi{10.1086/497536}
    
    \bibitem[{{Telloni} {et~al.}(2022){Telloni}, {Zank}, {Stangalini}, {Downs},
      {Liang}, {Nakanotani}, {Andretta}, {Antonucci}, {Sorriso-Valvo}, {Adhikari},
      {Zhao}, {Marino}, {Susino}, {Grimani}, {Fabi}, {D'Amicis}, {Perrone},
      {Bruno}, {Carbone}, {Mancuso}, {Romoli}, {Deppo}, {Fineschi}, {Heinzel},
      {Moses}, {Naletto}, {Nicolini}, {Spadaro}, {Teriaca}, {Frassati}, {Jerse},
      {Landini}, {Pancrazzi}, {Russano}, {Sasso}, {Biondo}, {Burtovoi}, {Capuano},
      {Casini}, {Casti}, {Chioetto}, {De Leo}, {Giarrusso}, {Liberatore},
      {Berghmans}, {Auch{\`e}re}, {Cuadrado}, {Chitta}, {Harra}, {Kraaikamp},
      {Long}, {Mandal}, {Parenti}, {Pelouze}, {Peter}, {Rodriguez}, {Sch{\"u}hle},
      {Schwanitz}, {Smith}, {Verbeeck}, \& {Zhukov}}]{Telloni_2022_ApJ}
    {Telloni}, D., {Zank}, G.~P., {Stangalini}, M., {et~al.} 2022, \apjl, 936, L25,
      \dodoi{10.3847/2041-8213/ac8104}
    
    \bibitem[{{Tenerani} {et~al.}(2020){Tenerani}, {Velli}, {Matteini},
      {R{\'e}ville}, {Shi}, {Bale}, {Kasper}, {Bonnell}, {Case}, {de Wit}, {Goetz},
      {Harvey}, {Klein}, {Korreck}, {Larson}, {Livi}, {MacDowall}, {Malaspina},
      {Pulupa}, {Stevens}, \& {Whittlesey}}]{Tenerani_2020_ApJS}
    {Tenerani}, A., {Velli}, M., {Matteini}, L., {et~al.} 2020, \apjs, 246, 32,
      \dodoi{10.3847/1538-4365/ab53e1}
    
    \bibitem[{{Thornton} \& {Parnell}(2011)}]{Thornton_2011_SolPhys}
    {Thornton}, L.~M., \& {Parnell}, C.~E. 2011, \solphys, 269, 13,
      \dodoi{10.1007/s11207-010-9656-7}
    
    \bibitem[{{Toriumi} \& {Airapetian}(2022)}]{Toriumi_2022_ApJ}
    {Toriumi}, S., \& {Airapetian}, V.~S. 2022, \apj, 927, 179,
      \dodoi{10.3847/1538-4357/ac5179}
    
    \bibitem[{{Toriumi} {et~al.}(2022){Toriumi}, {Airapetian}, {Namekata}, \&
      {Notsu}}]{Toriumi_2022_ApJS}
    {Toriumi}, S., {Airapetian}, V.~S., {Namekata}, K., \& {Notsu}, Y. 2022, \apjs,
      262, 46, \dodoi{10.3847/1538-4365/ac8b15}
    
    \bibitem[{{Truemper}(1982)}]{Truemper_1982_AdSpR}
    {Truemper}, J. 1982, Advances in Space Research, 2, 241,
      \dodoi{10.1016/0273-1177(82)90070-9}
    
    \bibitem[{{Tsuneta} {et~al.}(2008){Tsuneta}, {Ichimoto}, {Katsukawa}, {Lites},
      {Matsuzaki}, {Nagata}, {Orozco Su{\'a}rez}, {Shimizu}, {Shimojo}, {Shine},
      {Suematsu}, {Suzuki}, {Tarbell}, \& {Title}}]{Tsuneta_2008_ApJ}
    {Tsuneta}, S., {Ichimoto}, K., {Katsukawa}, Y., {et~al.} 2008, \apj, 688, 1374,
      \dodoi{10.1086/592226}
    
    \bibitem[{{Usmanov} {et~al.}(2018){Usmanov}, {Matthaeus}, {Goldstein}, \&
      {Chhiber}}]{Usmanov_2018_ApJ}
    {Usmanov}, A.~V., {Matthaeus}, W.~H., {Goldstein}, M.~L., \& {Chhiber}, R.
      2018, \apj, 865, 25, \dodoi{10.3847/1538-4357/aad687}
    
    \bibitem[{{Vaiana} {et~al.}(1981){Vaiana}, {Cassinelli}, {Fabbiano},
      {Giacconi}, {Golub}, {Gorenstein}, {Haisch}, {Harnden}, {Johnson}, {Linsky},
      {Maxson}, {Mewe}, {Rosner}, {Seward}, {Topka}, \& {Zwaan}}]{Vaiana_1981_ApJ}
    {Vaiana}, G.~S., {Cassinelli}, J.~P., {Fabbiano}, G., {et~al.} 1981, \apj, 245,
      163, \dodoi{10.1086/158797}
    
    \bibitem[{{van Ballegooijen} \&
      {Asgari-Targhi}(2016)}]{van_Ballegooijen_2016_ApJ}
    {van Ballegooijen}, A.~A., \& {Asgari-Targhi}, M. 2016, \apj, 821, 106,
      \dodoi{10.3847/0004-637X/821/2/106}
    
    \bibitem[{{van Ballegooijen} \&
      {Asgari-Targhi}(2017)}]{van_Ballegooijen_2017_ApJ}
    ---. 2017, \apj, 835, 10, \dodoi{10.3847/1538-4357/835/1/10}
    
    \bibitem[{{van Ballegooijen} {et~al.}(2011){van Ballegooijen}, {Asgari-Targhi},
      {Cranmer}, \& {DeLuca}}]{van_Ballegooijen_2011_ApJ}
    {van Ballegooijen}, A.~A., {Asgari-Targhi}, M., {Cranmer}, S.~R., \& {DeLuca},
      E.~E. 2011, \apj, 736, 3, \dodoi{10.1088/0004-637X/736/1/3}
    
    \bibitem[{{van der Holst} {et~al.}(2014){van der Holst}, {Sokolov}, {Meng},
      {Jin}, {Manchester}, {T{\'o}th}, \& {Gombosi}}]{van_der_Holst_2014_ApJ}
    {van der Holst}, B., {Sokolov}, I.~V., {Meng}, X., {et~al.} 2014, \apj, 782,
      81, \dodoi{10.1088/0004-637X/782/2/81}
    
    \bibitem[{{van der Walt} {et~al.}(2011){van der Walt}, {Colbert}, \&
      {Varoquaux}}]{van_der_Walt_2011_CSE}
    {van der Walt}, S., {Colbert}, S.~C., \& {Varoquaux}, G. 2011, Computing in
      Science and Engineering, 13, 22, \dodoi{10.1109/MCSE.2011.37}
    
    \bibitem[{{van Leer}(1979)}]{van_Leer_1979_JCP}
    {van Leer}, B. 1979, Journal of Computational Physics, 32, 101,
      \dodoi{10.1016/0021-9991(79)90145-1}
    
    \bibitem[{{van Saders} {et~al.}(2016){van Saders}, {Ceillier}, {Metcalfe},
      {Silva Aguirre}, {Pinsonneault}, {Garc{\'\i}a}, {Mathur}, \&
      {Davies}}]{van_Saders_2016_Nature}
    {van Saders}, J.~L., {Ceillier}, T., {Metcalfe}, T.~S., {et~al.} 2016, \nat,
      529, 181, \dodoi{10.1038/nature16168}
    
    \bibitem[{{van Saders} {et~al.}(2019){van Saders}, {Pinsonneault}, \&
      {Barbieri}}]{van_Saders_2019_ApJ}
    {van Saders}, J.~L., {Pinsonneault}, M.~H., \& {Barbieri}, M. 2019, \apj, 872,
      128, \dodoi{10.3847/1538-4357/aafafe}
    
    \bibitem[{{Velli}(1994)}]{Velli_1994_ApJ}
    {Velli}, M. 1994, \apjl, 432, L55, \dodoi{10.1086/187510}
    
    \bibitem[{{Verdini} {et~al.}(2019){Verdini}, {Grappin}, \&
      {Montagud-Camps}}]{Verdini_2019_SolPhys}
    {Verdini}, A., {Grappin}, R., \& {Montagud-Camps}, V. 2019, \solphys, 294, 65,
      \dodoi{10.1007/s11207-019-1458-y}
    
    \bibitem[{{Verdini} \& {Velli}(2007)}]{Verdini_2007_ApJ}
    {Verdini}, A., \& {Velli}, M. 2007, \apj, 662, 669, \dodoi{10.1086/510710}
    
    \bibitem[{{Verdini} {et~al.}(2010){Verdini}, {Velli}, {Matthaeus}, {Oughton},
      \& {Dmitruk}}]{Verdini_2010_ApJ}
    {Verdini}, A., {Velli}, M., {Matthaeus}, W.~H., {Oughton}, S., \& {Dmitruk}, P.
      2010, \apjl, 708, L116, \dodoi{10.1088/2041-8205/708/2/L116}
    
    \bibitem[{{Veronig} {et~al.}(2021){Veronig}, {Odert}, {Leitzinger}, {Dissauer},
      {Fleck}, \& {Hudson}}]{Veronig_2021_NatAs}
    {Veronig}, A.~M., {Odert}, P., {Leitzinger}, M., {et~al.} 2021, Nature
      Astronomy, 5, 697, \dodoi{10.1038/s41550-021-01345-9}
    
    \bibitem[{{Vidotto}(2021)}]{Vidotto_2021_LRSP}
    {Vidotto}, A.~A. 2021, Living Reviews in Solar Physics, 18, 3,
      \dodoi{10.1007/s41116-021-00029-w}
    
    \bibitem[{{Vidotto} \& {Bourrier}(2017)}]{Vidotto_2017_MNRAS}
    {Vidotto}, A.~A., \& {Bourrier}, V. 2017, \mnras, 470, 4026,
      \dodoi{10.1093/mnras/stx1543}
    
    \bibitem[{{Vidotto} \& {Cleary}(2020)}]{Vidotto_2020_MNRAS}
    {Vidotto}, A.~A., \& {Cleary}, A. 2020, \mnras, 494, 2417,
      \dodoi{10.1093/mnras/staa852}
    
    \bibitem[{{Vidotto} \& {Donati}(2017)}]{Vidotto_2017_AA}
    {Vidotto}, A.~A., \& {Donati}, J.~F. 2017, \aap, 602, A39,
      \dodoi{10.1051/0004-6361/201629700}
    
    \bibitem[{{Vidotto} {et~al.}(2014){Vidotto}, {Gregory}, {Jardine}, {Donati},
      {Petit}, {Morin}, {Folsom}, {Bouvier}, {Cameron}, {Hussain}, {Marsden},
      {Waite}, {Fares}, {Jeffers}, \& {do Nascimento}}]{Vidotto_2014_MNRAS}
    {Vidotto}, A.~A., {Gregory}, S.~G., {Jardine}, M., {et~al.} 2014, \mnras, 441,
      2361, \dodoi{10.1093/mnras/stu728}
    
    \bibitem[{{V{\"o}gler} {et~al.}(2005){V{\"o}gler}, {Shelyag}, {Sch{\"u}ssler},
      {Cattaneo}, {Emonet}, \& {Linde}}]{Vogler_2005_AA}
    {V{\"o}gler}, A., {Shelyag}, S., {Sch{\"u}ssler}, M., {et~al.} 2005, \aap, 429,
      335, \dodoi{10.1051/0004-6361:20041507}
    
    \bibitem[{{Wang}(1994)}]{Wang_1994_ApJ}
    {Wang}, Y.~M. 1994, \apjl, 435, L153, \dodoi{10.1086/187617}
    
    \bibitem[{{Wang}(2020)}]{Wang_2020_ApJ}
    ---. 2020, \apj, 904, 199, \dodoi{10.3847/1538-4357/abbda6}
    
    \bibitem[{{Wang}(2022)}]{Wang_2022_SolPhys}
    ---. 2022, \solphys, 297, 129, \dodoi{10.1007/s11207-022-02060-y}
    
    \bibitem[{{Wang} \& {Sheeley}(1990)}]{Wang_1990_ApJ}
    {Wang}, Y.~M., \& {Sheeley}, N.~R., J. 1990, \apj, 355, 726,
      \dodoi{10.1086/168805}
    
    \bibitem[{{Wang} \& {Sheeley}(2002)}]{Wang_2002_JGR}
    {Wang}, Y.~M., \& {Sheeley}, N.~R. 2002, Journal of Geophysical Research (Space
      Physics), 107, 1302, \dodoi{10.1029/2001JA000500}
    
    \bibitem[{{Wargelin} \& {Drake}(2002)}]{Wargelin_2002_ApJ}
    {Wargelin}, B.~J., \& {Drake}, J.~J. 2002, \apj, 578, 503,
      \dodoi{10.1086/342270}
    
    \bibitem[{{Weber} \& {Davis}(1967)}]{Weber_1967_ApJ}
    {Weber}, E.~J., \& {Davis}, Leverett, J. 1967, \apj, 148, 217,
      \dodoi{10.1086/149138}
    
    \bibitem[{{Wiegelmann} \& {Solanki}(2004)}]{Wiegelmann_2004_SolPhys}
    {Wiegelmann}, T., \& {Solanki}, S.~K. 2004, \solphys, 225, 227,
      \dodoi{10.1007/s11207-004-3747-2}
    
    \bibitem[{{Wood} {et~al.}(2014){Wood}, {M{\"u}ller}, {Redfield}, \&
      {Edelman}}]{Wood_2014_ApJ}
    {Wood}, B.~E., {M{\"u}ller}, H.-R., {Redfield}, S., \& {Edelman}, E. 2014,
      \apjl, 781, L33, \dodoi{10.1088/2041-8205/781/2/L33}
    
    \bibitem[{{Wood} {et~al.}(2002){Wood}, {M{\"u}ller}, {Zank}, \&
      {Linsky}}]{Wood_2002_ApJ}
    {Wood}, B.~E., {M{\"u}ller}, H.-R., {Zank}, G.~P., \& {Linsky}, J.~L. 2002,
      \apj, 574, 412, \dodoi{10.1086/340797}
    
    \bibitem[{{Wood} {et~al.}(2005){Wood}, {M{\"u}ller}, {Zank}, {Linsky}, \&
      {Redfield}}]{Wood_2005_ApJ}
    {Wood}, B.~E., {M{\"u}ller}, H.~R., {Zank}, G.~P., {Linsky}, J.~L., \&
      {Redfield}, S. 2005, \apjl, 628, L143, \dodoi{10.1086/432716}
    
    \bibitem[{{Wood} {et~al.}(2021){Wood}, {M{\"u}ller}, {Redfield}, {Konow},
      {Vannier}, {Linsky}, {Youngblood}, {Vidotto}, {Jardine},
      {Alvarado-G{\'o}mez}, \& {Drake}}]{Wood_2021_ApJ}
    {Wood}, B.~E., {M{\"u}ller}, H.-R., {Redfield}, S., {et~al.} 2021, \apj, 915,
      37, \dodoi{10.3847/1538-4357/abfda5}
    
    \bibitem[{{Woods} \& {Rottman}(2005)}]{Woods_2005_SolPhys_XPS_SORCE}
    {Woods}, T.~N., \& {Rottman}, G. 2005, \solphys, 230, 375,
      \dodoi{10.1007/s11207-005-2555-7}
    
    \bibitem[{{Woods} {et~al.}(2005){Woods}, {Rottman}, \&
      {Vest}}]{Woods_2005_SolPhys_XPS_overview_calibration}
    {Woods}, T.~N., {Rottman}, G., \& {Vest}, R. 2005, \solphys, 230, 345,
      \dodoi{10.1007/s11207-005-4119-2}
    
    \bibitem[{{Wright} \& {Drake}(2016)}]{Wright_2016_Nature}
    {Wright}, N.~J., \& {Drake}, J.~J. 2016, \nat, 535, 526,
      \dodoi{10.1038/nature18638}
    
    \bibitem[{{Wyper} {et~al.}(2022){Wyper}, {DeVore}, {Antiochos}, {Pontin},
      {Higginson}, {Scott}, {Masson}, \& {Pelegrin-Frachon}}]{Wyper_2022_ApJ}
    {Wyper}, P.~F., {DeVore}, C.~R., {Antiochos}, S.~K., {et~al.} 2022, \apjl, 941,
      L29, \dodoi{10.3847/2041-8213/aca8ae}
    
    \bibitem[{{Yokoyama} \& {Shibata}(2001)}]{Yokoyama_2001_ApJ}
    {Yokoyama}, T., \& {Shibata}, K. 2001, \apj, 549, 1160, \dodoi{10.1086/319440}
    
    \bibitem[{{Yoshida} {et~al.}(2023){Yoshida}, {Shimizu}, \&
      {Toriumi}}]{Yoshida_2023_ApJ}
    {Yoshida}, M., {Shimizu}, T., \& {Toriumi}, S. 2023, \apj, 950, 156,
      \dodoi{10.3847/1538-4357/acd053}
    
    \bibitem[{{Zank} {et~al.}(2018){Zank}, {Adhikari}, {Hunana}, {Tiwari}, {Moore},
      {Shiota}, {Bruno}, \& {Telloni}}]{Zank_2018_ApJ}
    {Zank}, G.~P., {Adhikari}, L., {Hunana}, P., {et~al.} 2018, \apj, 854, 32,
      \dodoi{10.3847/1538-4357/aaa763}
    
    \bibitem[{{Zank} {et~al.}(2020){Zank}, {Nakanotani}, {Zhao}, {Adhikari}, \&
      {Kasper}}]{Zank_2020_ApJ}
    {Zank}, G.~P., {Nakanotani}, M., {Zhao}, L.~L., {Adhikari}, L., \& {Kasper}, J.
      2020, \apj, 903, 1, \dodoi{10.3847/1538-4357/abb828}
    
    \bibitem[{{Zhuleku} {et~al.}(2020){Zhuleku}, {Warnecke}, \&
      {Peter}}]{Zhuleku_2020_AA}
    {Zhuleku}, J., {Warnecke}, J., \& {Peter}, H. 2020, \aap, 640, A119,
      \dodoi{10.1051/0004-6361/202038022}
    
    \end{thebibliography}

\end{document}